\documentclass[lettersize,journal]{IEEEtran}

\usepackage{amsmath,amsfonts,amssymb}
\usepackage{algorithmic}
\usepackage{array}
\usepackage{subfig}
\usepackage{textcomp}
\usepackage{stfloats}
\usepackage{url}
\usepackage{verbatim}
\usepackage{graphicx}
\def\BibTeX{{\rm B\kern-.05em{\sc i\kern-.025em b}\kern-.08em
    T\kern-.1667em\lower.7ex\hbox{E}\kern-.125emX}}
\usepackage{balance}

\usepackage{amsthm}
\usepackage{graphicx}
\usepackage{import}
\usepackage{calc}
\usepackage{xcolor} 
\usepackage{todonotes}
\usepackage[linewidth=1pt]{mdframed} 
\usepackage{cuted} 
\usepackage{multirow}

\newcommand*\subtxt[1]{_{\mathrm{#1}}}
\DeclareRobustCommand\_{\ifmmode\expandafter\subtxt\else\textunderscore\fi}

\newcommand{\rmj}{\mathrm{j}}
\newcommand{\mrm}{\mathrm}
\newcommand{\mbf}{\mathbf}

\newtheorem{rem}{Remark}

\graphicspath{{figs/}}

\begin{document}
\title{On the Analysis and Synthesis of Wind Turbine Side-Side Tower Load Control via Demodulation}
\author{
Atindriyo K. Pamososuryo, Sebastiaan P. Mulders, Riccardo Ferrari, and Jan-Willem van Wingerden
\thanks{
Atindriyo K. Pamososuryo, Sebastiaan P. Mulders, Riccardo Ferrari, and Jan-Willem van Wingerden are with the Delft Center for Systems and Control, Delft University of Technology, Mekelweg 2, 2628 CD Delft, the Netherlands (e-mail: {\{A.K.Pamososuryo, S.P.Mulders, R.Ferrari, J.W.vanWingerden\}@tudelft.nl}).}}


\maketitle

\begin{abstract}
As wind turbine power capacities continue to rise, taller and more flexible tower designs are needed for support. 
These designs often have the tower's natural frequency in the turbine's operating regime, increasing the risk of resonance excitation and fatigue damage.
Advanced load-reducing control methods are needed to enable flexible tower designs that consider the complex dynamics of flexible turbine towers during partial-load operation.
This paper proposes a novel modulation-demodulation control (MDC) strategy for side-side tower load reduction driven by the varying speed of the turbine.
The MDC method demodulates the periodic content at the once-per-revolution (1P) frequency in the tower motion measurements into two orthogonal channels.
The proposed scheme extends the conventional tower controller by augmentation of the MDC contribution to the generator torque signal.
A linear analysis framework into the multivariable system in the demodulated domain reveals varying degrees of coupling at different rotational speeds and a gain sign flip.
As a solution, a decoupling strategy has been developed, which simplifies the controller design process and allows for a straightforward (but highly effective) diagonal linear time-invariant controller design.
The high-fidelity OpenFAST wind turbine software evaluates the proposed controller scheme, demonstrating effective reduction of the 1P periodic loading and the tower's natural frequency excitation in the side-side tower motion.
\end{abstract}

\begin{IEEEkeywords}
wind turbine fatigue reduction, side-side tower load control, periodic load cancellation, modulation-demodulation control
\end{IEEEkeywords}


\section{Introduction}\label{sec:1} 
\noindent
\IEEEPARstart{T}{he} improvements in the cost-effectiveness of wind turbines can be traced back to the adoption of the upscaling strategy, in which the sizes of the turbine components are made larger~\cite{Veers2019}.
Turbine rotors are made larger and can capture more energy by the increased swept area and towers are built taller to support such larger rotors. 
Moreover, at higher altitudes, the wind energy resource is of higher quality as the influence of surface friction is less prominent, resulting in more power production by a single machine.

Conventional tower upscaling, however, is not desirable as merely increasing tower heights and diameters while keeping the same wall thickness results in much heavier and more expensive structures.
In addition, transportability constraints, e.g., on tower base diameter, limit the size of which the towers can be designed, especially for onshore installations~\cite{Dykes2018TallTowers}.
While satisfying the transportability requirements, reducing the tower wall thickness is a compelling solution to lower the needed amount of mass and thus manufacturing costs of tall towers.
In contrast to conventional \textit{soft-stiff} tower designs, resulting \textit{soft-soft} designs lower the tower's (first) natural frequency into the turbine operational range.
Consequently, the risk of resonance by excitation of the time-varying rotor speed, also known as the once-per-revolution (1P) frequency, is thus becoming ever greater.
Rotor imbalance could even further exacerbate this effect~\cite{Ramlau2009,Saathoff2021}.
Tower resonance excitation is even more concerning for the side-side tower oscillations than for the fore-aft due to the negligible contribution of the aerodynamic damping in the formerly mentioned direction~\cite{Burton2011}.
So, reliable and advanced control solutions capable of fatigue load mitigation are of utmost importance to improve the viability of soft-soft tower designs.

Different control implementations have been made available in the literature for tower periodic load control and are generally classified as passive and active methods.
Passive methods prevent prolonged turbine operation near the tower resonance frequency, usually by decreasing or increasing generator torque demand to accelerate or decelerate the rotor, depending on whether its speed is above or below the resonance frequency.
This method is often referred to as frequency skipping by speed exclusion zones.
Bossanyi introduced the approach for avoiding tower resonance by blade passing frequency at three-times-per-revolution (3P) for three-bladed wind turbines~\cite{Bossanyi2003}.
Licari et al. studied the effects of the speed exclusion zone's width tuning for 1P excitation in terms of load reduction and power quality~\cite{Licari2013}.
Smilden and S{\o}rensen later adopted this algorithm for preventing resonance of fore-aft tower motion by the 3P thrust oscillations~\cite{Smilden2016}.
However, such conventional implementations are non-trivial due to the intricate logic that needs to be incorporated, resulting in not knowing whether the control solution is dynamically optimal.
Therefore, a state-of-the-art quasi-linear parameter varying model predictive control (qLPV-MPC) method was developed to tackle the shortcomings and challenges of conventional methods~\cite{Mulders2020}.

Active control methods, on the other hand, feed tower measurements into a controller to generate counteracting forces through provided actuators so as to dampen the tower vibration~\cite{Bossanyi2003,vanDerHooft2003}.
The controller, typically an integrator when acceleration is measured, is designed to increase tower damping.
Depending on whether the fore-aft~\cite{Bossanyi2003,vanDerHooft2003,Wright2008} or side-side direction~\cite{vanDerHooft2003,Zhang2014} is targeted for damping, respectively, the collective pitch or generator torque demand is utilized as the control input.
Such a conventional approach was originally devised to reduce tower vibration at its natural frequency.
Nonetheless, a tower load controller specifically aiming at the time-varying 1P periodic loading has received little to no attention in the literature and would be an attractive complement to the conventional method.

This paper extends the conventional control method by a modulation-demodulation setup to further improve side-side tower load reduction performance. 
Such a controller design additionally provides a reduction of the rotational-speed-driven load and allows for comprehensive system analysis and controller synthesis.
To the best of the authors' knowledge, such an approach has not yet received any attention in the wind turbine control literature.

In the control engineering field, an approach known as the modulated-demodulated control (MDC) is considered an effective solution to periodic disturbance cancellation problems~\cite{Bodson1994,Byl2005,Lau2007}.
MDC is able to adapt its control input's frequency to reject a time-varying disturbance frequency and can handle the changes in the dynamics of the plant due to the variation of the disturbance frequency.
A large body of literature has been dedicated to further studying MDC's potential, which includes applications in diverse fields.
For instance, \cite{Byl2005} focuses on MDC tuning from the perspective of the frequency-domain loop-shaping method for the application of a diamond turning machine.
The work of~\cite{Lau2005} analyses the feedback limitations of MDC by poles, zeros, and delays investigation.
In~\cite{Lau2007}, vibration control of flexible piezoelectric structures by MDC was conducted.
Other applications include digital data storage system~\cite{Sacks1995}, helicopter~\cite{Ariyur1999}, flexible web winding system~\cite{Xu2002}, and tape system~\cite{Zhong2005}, just to mention a few.

The MDC control method bears similarities with a highly-anticipated and industrially applied periodic blade load alleviation technique known as individual pitch control (IPC)~\cite{Bossanyi2003b}.
For most IPC implementations, the Coleman transformation~\cite{Bir2008} is employed to project measured individual blade moments containing periodic content from the rotating frame into a static nonrotating frame.
The scheme can be thought of as a modulation-demodulation framework where structural analysis and controller design are simplified in the nonrotating domain. 
However, it has been known that larger and more flexible rotor structures create severe coupling of the considered multivariable system.
Therefore, for single-input single-output (SISO) controller designs to be justified in the transformed domain, decoupling strategies must be taken into account~\cite{Unguran2019,Mulders2019}.
In MDC, identical and rather simple diagonal SISO controllers can also be designed onto low-frequency, orthogonal \textit{quadrature} and \textit{in-phase} channels resulting from the \textit{demodulation} of the plant's measurements.
This operation is similar to the forward Coleman transformation in the conventional IPC.
The computed control actions on these orthogonal channels are then converted into the actual usable input at the disturbance frequency by the \textit{modulation} operation, similar to the reverse Coleman transformation.
Nevertheless, despite its demonstrated effectiveness in wide applications, little attention has been paid to adapting MDC for mitigating periodic loading affecting wind turbine side-side tower motion.

This paper focuses on the development of MDC for the rejection of 1P periodic loading on wind turbine side-side tower motion.
The proposed MDC results in a periodic generator torque control input with time-varying 1P frequency, which, given the measurements of the rotor speed, is able to track the disturbance's frequency.
However, frequency-domain analysis of the demodulated system shows that the quadrature and in-phase channels are not fully decoupled.
Moreover, changing rotational speed induces a gain sign flip, which may cause instability in the closed-loop operation.
Therefore, a decoupling strategy by phase offset inclusion, similar to that in the conventional IPC~\cite{Mulders2019}, is developed to arrive at fully decoupled quadrature and in-phase channels and simultaneously remove the gain sign flip.

The contribution of this work is four-fold:
\begin{enumerate}
    \item{Formulating MDC for the mitigation of periodic load affecting wind turbine side-side tower motion induced by the time-varying 1P rotor excitation;}
    \item{Providing frequency domain frameworks for the analysis of the system coupling and controller behavior in their (de)modulated representations;}
    \item{Decoupling the multivariable system and correcting the gain sign flip by the inclusion of a phase offset, as well as illustrating the offset's influence on the controller;}
    \item{Showcasing the performance of MDC in both simplified and high-fidelity computer-aided wind turbine simulation environments along with a conventional active tower damper.}
\end{enumerate}

The remainder of this paper is structured as follows.
Section~\ref{sec:2} describes the nominal wind turbine dynamics and conventional tower damping controller.
In Section~\ref{sec:3}, the derivation of the proposed MDC framework in the frequency domain is presented.
Section~\ref{sec:4} elaborates on controller and system analysis in the MDC framework, in which the cross-coupling phenomenon and gain sign flip in the quadrature and in-phase MDC channels are discussed.
Then, in Section~\ref{sec:5}, the phase offset inclusion for the channel decoupling and gain sign flip correction on the tower dynamics, as well as influence on the controller, is explained.
In Section~\ref{sec:6}, the effectiveness of the proposed controller is demonstrated using low- and high-fidelity simulations.
Concluding remarks are drawn in Section~\ref{sec:7}.

\section{Wind Turbine Dynamics and Conventional Tower Damping Controller}\label{sec:2} 
\noindent
To form the basis for controller design and analysis in this paper, wind turbine aerodynamic and tower models are derived in Section~\ref{sec:2_dynamics}.
As the goal of this paper is to augment the proposed controller to that of the conventional active tower damping controller, the latter's design is discussed in Section~\ref{sec:2_baseline_ctrl} and is combined with the tower dynamics. 
Section~\ref{sec:2_freq_domain} derives the frequency domain representation of the combined tower dynamics for later uses in the MDC framework.

\subsection{Tower and aerodynamic models}\label{sec:2_dynamics}
\noindent
The model used for analysis and synthesis considered in this study consists of side-side tower dynamics and rotor aerodynamics.
The tower dynamics are approximated as a second-order system, representing the first structural mode as follows:
\begin{multline}\label{eq:dynamics_tower}
    \ddot{x}(t) = \frac{1}{m} \left(- d \dot{x}(t) - k x(t) + F\_{sd}(t) \right.\\
    \left. + s\_{f} (T\_{g}(t) + \Delta T\_{g,damp}(t) + \Delta T\_{g}(t)) \right) \,,
\end{multline}
where $m$ denotes the tower modal mass, $d$ its damping, and $k$ the modal stiffness.
The notation $t$ indicates a time-domain signal, which, for the sake of brevity, is omitted in the text unless necessary.
Tower-top acceleration, velocity, and displacement are represented by $\ddot{x}$, $\dot{x}$, and $x$, respectively.
The motion of the tower is affected by generator torque activities $T\_{g}$, $\Delta T\_{g,damp}$, and $\Delta T\_{g}$ through the generator stator reaction, all of which are the considered control actions in this paper.
The torque $T\_{g}$ is used mainly in power production, whereas the additive torques $\Delta T\_{g,damp}$ and $\Delta T\_{g}$ are utilized to respectively increase the effective tower damping and mitigate the periodic loading $F\_{sd}$.
The factor $s\_{f} = 1.5/H$, with $H$ being the tower height, is the ratio between the rotational and translational displacements of the tower motion under the assumption that the tower is a prismatic beam~\cite{Selvam2009a}.

\begin{figure}[t]
    \centering
    \def\svgscale{0.8}
\begingroup%
  \makeatletter%
  \providecommand\color[2][]{%
    \errmessage{(Inkscape) Color is used for the text in Inkscape, but the package 'color.sty' is not loaded}%
    \renewcommand\color[2][]{}%
  }%
  \providecommand\transparent[1]{%
    \errmessage{(Inkscape) Transparency is used (non-zero) for the text in Inkscape, but the package 'transparent.sty' is not loaded}%
    \renewcommand\transparent[1]{}%
  }%
  \providecommand\rotatebox[2]{#2}%
  \newcommand*\fsize{\dimexpr\f@size pt\relax}%
  \newcommand*\lineheight[1]{\fontsize{\fsize}{#1\fsize}\selectfont}%
  \ifx\svgwidth\undefined%
    \setlength{\unitlength}{286.40720985bp}%
    \ifx\svgscale\undefined%
      \relax%
    \else%
      \setlength{\unitlength}{\unitlength * \real{\svgscale}}%
    \fi%
  \else%
    \setlength{\unitlength}{\svgwidth}%
  \fi%
  \global\let\svgwidth\undefined%
  \global\let\svgscale\undefined%
  \makeatother%
  \begin{picture}(1,1.12449198)%
    \lineheight{1}%
    \setlength\tabcolsep{0pt}%
    \put(0,0){\includegraphics[width=\unitlength,page=1]{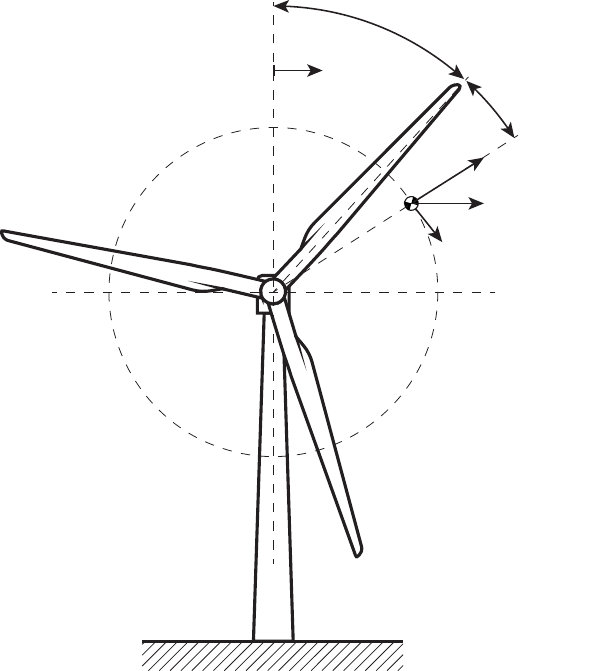}}%
    \put(0.83011346,0.84866634){\color[rgb]{0.1372549,0.12156863,0.1254902}\makebox(0,0)[lt]{\lineheight{1.25}\smash{\begin{tabular}[t]{l}$a_\mathrm{sd}$\end{tabular}}}}%
    \put(0.74152515,0.69751){\color[rgb]{0.1372549,0.12156863,0.1254902}\makebox(0,0)[lt]{\lineheight{1.25}\smash{\begin{tabular}[t]{l}$v_\mathrm{t}$\end{tabular}}}}%
    \put(0.47222596,1.02778064){\color[rgb]{0.1372549,0.12156863,0.1254902}\makebox(0,0)[lt]{\lineheight{1.25}\smash{\begin{tabular}[t]{l}$x$\end{tabular}}}}%
    \put(0.63362299,1.09271291){\color[rgb]{0.1372549,0.12156863,0.1254902}\makebox(0,0)[lt]{\lineheight{1.25}\smash{\begin{tabular}[t]{l}$\psi$\end{tabular}}}}%
    \put(0.82732259,0.94592403){\color[rgb]{0.1372549,0.12156863,0.1254902}\makebox(0,0)[lt]{\lineheight{1.25}\smash{\begin{tabular}[t]{l}$\phi_\mathrm{sd}$\end{tabular}}}}%
    \put(0.81713729,0.77925676){\color[rgb]{0.1372549,0.12156863,0.1254902}\makebox(0,0)[lt]{\lineheight{1.25}\smash{\begin{tabular}[t]{l}$F_\mathrm{sd}$\end{tabular}}}}%
  \end{picture}%
\endgroup%

    \caption{
    A wind turbine is excited at the side-side direction by a periodic load due to the rotor imbalance at the 1P frequency ${F_\text{sd}(t) = a\_{sd} \cos{(\psi(t) + \phi\_{sd})}}$, with the azimuth $\psi(t)=\omega\_{r}(t)t$. 
    The tangential speed of the periodic load is indicated by $v\_{t}(t)$, and $x(t)$ denotes tower top displacement in the horizontal direction.
    }
    \label{fig:tower_schematic}
\end{figure}
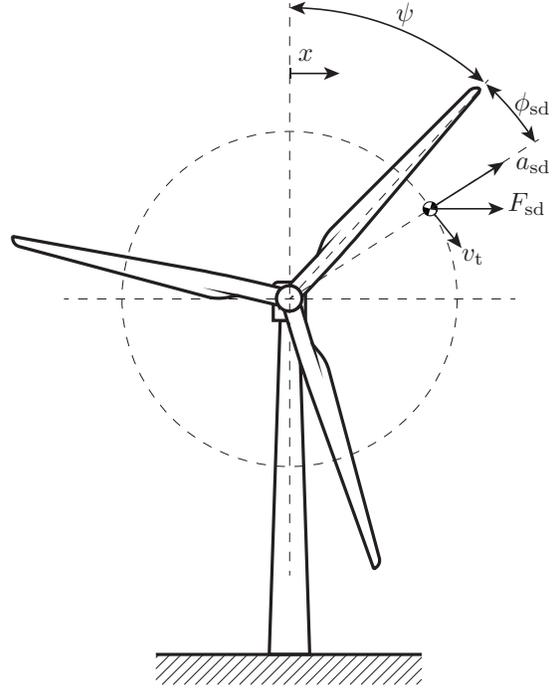

The periodic loading aforementioned may develop on the rotor due to, e.g., a mass or aerodynamic imbalance and transferred to the fixed structure, which is modeled as the following sinusoidal force~\cite{Pamososuryo2022}
\begin{equation}\label{eq:force_side_nominal_old}
    F\_{sd}(t) = a\_{sd} \cos{(\omega\_{r}(t) t + \phi\_{sd})} \,,
\end{equation}
where its amplitude and phase offset are denoted as $a\_{sd}$ and $\phi\_{sd}$, respectively.
Figure~\ref{fig:tower_schematic} illustrates how this force affects the tower.

The frequency of $F\_{sd}$ varies in time as the rotor speed $\omega\_{r}$ changes according to the following rotor aerodynamics, resembling a first-order rotational mass system
\begin{equation}\label{eq:dynamics_drivetrain}
     \dot{\omega}\_{r}(t) = \frac{1}{J\_{r}} T\_{a}(t) - \frac{G}{J\_{r}} (T\_{g}(t) + \Delta T\_{g,damp}(t) + \Delta T\_{g}(t)) \,.
\end{equation}
In the above equation, $J\_{r}$ represents the equivalent inertia at the low-speed-shaft (LSS) side and $G$ is the gearbox ratio.
The aerodynamic torque is given by
\begin{equation*}\label{eq:torque_aero}
    T\_{a}(t) = \frac{1}{2\omega\_{r}(t)} \rho\_{a} \pi R^2 C\_{p}\bigl(\omega\_{r}(t),v(t),\beta(t)\bigr) v(t)^3 \, ,
\end{equation*}
with the air density denoted by $\rho\_{a}$, rotor radius $R$, and aerodynamic power coefficient $C\_{p}$, being a function of $\omega\_{r}$, wind speed $v$ and pitch angle $\beta$.
To achieve maximum power extraction at the below-rated operating regime, as considered in this work, the so-called $K\omega\_{r}^2$ control law~\cite{Bossanyi2000} is employed for the torque controller
\begin{equation}\label{eq:Komega2}
    T\_{g}(t) = \underbrace{\frac{1}{2 {\lambda^\star}^3} \rho\_{a} \pi R^5 C\_{p}^\star}_{K} \omega\_{r}(t)^2 \,,
\end{equation}
where $K$ is the optimal gain, $\lambda^\star$ as the design tip-speed ratio corresponding to the optimal power coefficient $C\_{p}^\star$ at fine pitch position.
This work employs only a simple maximum power tracking controller, as load mitigation is the main focus.
However, a more advanced method is available in the literature for the interested reader, e.g.~\cite{Brandetti2023}.

\subsection{Conventional Active Tower Damping Controller}\label{sec:2_baseline_ctrl}
\noindent
With the wind turbine model at hand, a side-side tower damping controller is added to obtain the nominal system considered in the remainder of this paper.
The wind turbine tower dynamics~\eqref{eq:dynamics_tower} commonly possess only negligible damping $d$ such that additional damping is required to mitigate fatigue loads at the tower's natural frequency.
Conventionally, for the side-side direction, the extra damping is created by additional generator torque demand, being negatively proportional to the tower-top velocity (also taking into account $s\_{f}$) as follows~\cite{vanDerHooft2003}
\begin{align}\label{eq:damping_add}
    \Delta T\_{g,damp}(t) = -K\_{conv} \dot{x}(t) \,,
\end{align}
with
\begin{equation}
    K\_{conv} = \frac{d\_{add}}{s\_{f}} \,,
\end{equation}
as a constant gain where $d\_{add}$ is the additional, desired modal damping.
Note that since $\ddot{x}$ is often measurable, it is necessary to perform integration of this signal to obtain $\dot{x}$; thus, this controller is essentially an integral controller.

The increase in the effective modal damping coefficient is now evident by substituting $\Delta T\_{g,damp}$ in~\eqref{eq:dynamics_tower} with~\eqref{eq:damping_add}, such that the tower dynamics are rendered into
\begin{multline}\label{eq:dynamics_tower_damped}
    \ddot{x}(t) = \frac{1}{m}\left(-d\_{eff} \dot{x}(t) - k x(t) + F\_{sd}(t) \right.\\
    \left. + s\_{f} (T\_{g}(t) + \Delta T\_{g}(t)) \right)\,,
\end{multline}
with $d\_{eff} = d+d\_{add}$ being the effective damping.
However, the conventional tower damper does not focus on the mitigation of 1P periodic loading posed by $F\_{sd}$.
Later on in this paper, the development of MDC for alleviating this rotor-speed-driven load is discussed further as an extension to this conventional controller, making use of~\eqref{eq:dynamics_tower_damped}.

\subsection{Frequency Domain Representation}\label{sec:2_freq_domain}
\noindent
In the following sections, MDC design and analysis are done in the frequency domain.
This requires the dynamics~\eqref{eq:dynamics_tower_damped} to be expressed in this domain as well, where the transfer from $\Delta T\_{g}$ to $\dot{x}$ is considered.
The Laplace transformation of the tower dynamics gives the following transfer function
\begin{equation}\label{eq:Gs}
    \mrm{G}(s) = \frac{\dot{\mrm{X}}(s)}{\Delta \mrm{T}\_{g}(s)} = \frac{s\_{f} s}{m s^2 + d\_{eff}s + k} \,,
\end{equation}
where $s$ is the Laplace operator.
The notations $\dot{\mrm{X}}(s)$ and $\Delta \mrm{T}\_{g}(s)$ are the tower-top velocity and additive generator torque in their frequency domain representation.
Also useful is to define $\mrm{N}(s)$ and $\mrm{D}(s)$ to denote the numerator and denominator of $\mrm{G}(s)$.

\section{Modulated-Demodulated Control Scheme}\label{sec:3} 
\noindent
This section provides an elaboration on the MDC architecture, which is based on the approach proposed by~\cite{Lau2007}.
The MDC architecture and accompanying analysis methods are utilized to provide a control-oriented analysis of the controller and wind turbine system.
As a prerequisite for subsequent derivations, the following Laplace transforms of signal modulations at the disturbance frequency $\omega\_{r}$ are defined
\begin{subequations}\label{eq:laplace_cossin}
\begin{align}
    \mathcal{L}\left\{ r(t) \cos({\omega\_{r}t}) \right\} &= \frac{1}{2} \left( \mrm{R}(s_{-}) + \mrm{R}(s_{+}) \right)      \,, \label{eq:laplace_cos}\\
    \mathcal{L}\left\{ r(t) \sin({\omega\_{r}t}) \right\} &= -\frac{\rmj }{2} \left( \mrm{R}(s_{-}) - \mrm{R}(s_{+}) \right) \,, \label{eq:laplace_sin}
\end{align}
\end{subequations}
with $r(t)$ as an arbitrary time-domain signal and $\mrm{R}(s)$ its Laplace-transformed analogue. 
For this linear analysis, $\omega\_{r}$ is assumed to be constant over a single period.
The notation ${s_{\pm} = s \pm \rmj \omega\_{r}}$ is introduced to indicate $\omega\_{r}$-shifted frequency content.
Another useful relation for the derivations that follow is the following Euler's formula
\begin{equation}\label{eq:euler_formula}
    e^{\rmj \phi}=\cos{(\phi)}+ \rmj \sin{(\phi)} \,,
\end{equation}
with $\phi$ as an arbitrary angle.

\begin{figure*}[!t]
    \centering
    \includegraphics[width=0.95\linewidth]{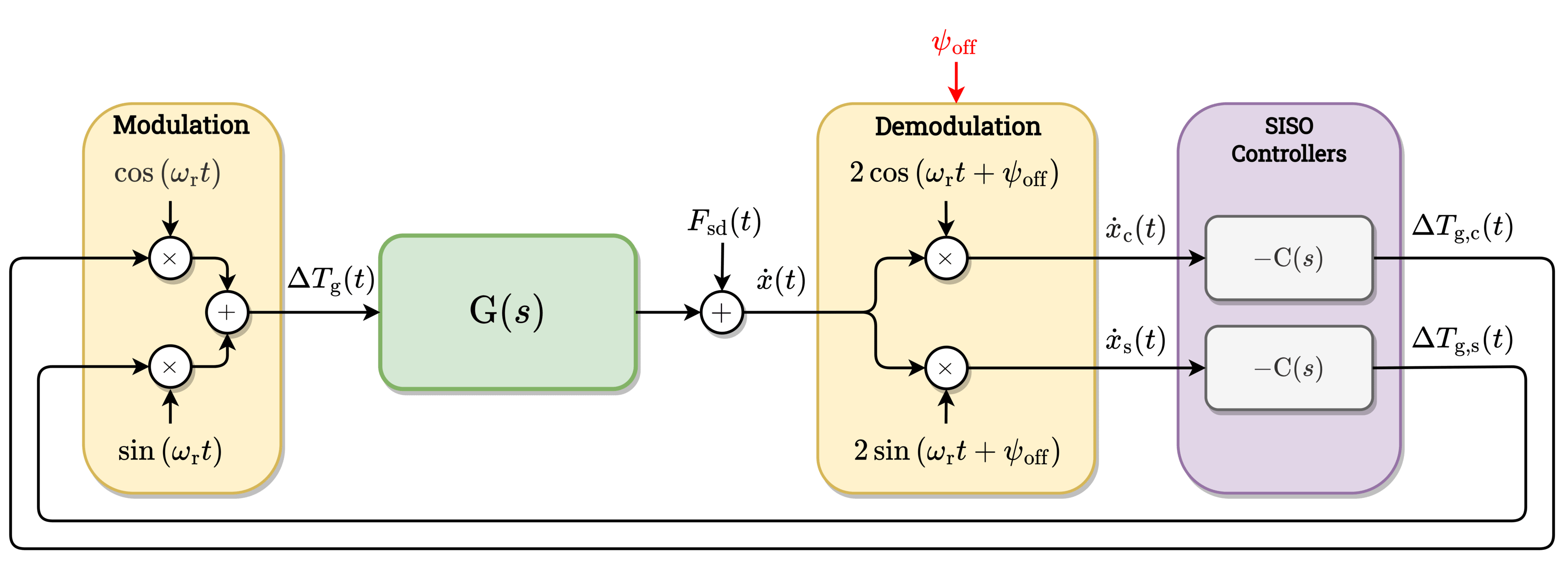}
    \caption{
    The modulated-demodulated control scheme for the cancellation of a side-side periodic load $F\_{sd} = a\_{sd} \cos{(\omega\_{r}t+\phi\_{sd})}$ affecting wind turbine tower $\mrm{G}(s)$.
    The demodulation operation is driven by the disturbance frequency $\omega\_{r}$ and creates a separation of the output signal $\dot{x}$ into quadrature and in-phase signals, $\dot{x}\_{c}$ and $\dot{x}\_{s}$, respectively.
    On these channels, two identical SISO controllers $\mrm{C}(s) \mbf{I}_{2\times2}$ are designed, generating the control inputs $\Delta T\_{g,c}$ and $\Delta T\_{g,s}$.
    Finally, modulation at $\omega\_{r}$ combines these two control inputs into a single signal $\Delta T\_{g}$ that is fed into $\mrm{G}(s)$ to alleviate the periodic loading.
    The phase offset $\psi\_{off}$ can be added to the demodulator to influence the system's behavior, such as channel decoupling.
    Note that the negative sign preceding $\mrm{C}(s)$ indicates the negative feedback convention used in the framework.
    }
    \label{fig:ModulatedDemodulatedWT}
\end{figure*}

The MDC methodology follows the depiction in Fig.~\ref{fig:ModulatedDemodulatedWT}, where modulation and demodulation involve signal multiplications with trigonometric functions at the disturbance frequency $\omega\_{r}$.
In contrast to the work of~\cite{Lau2007}, any filtering in the demodulation stage is omitted and is relocated to the controller block/stage for the sake of generalization.

Demodulation is the first stage of the MDC scheme, where $\dot{x}$, being the output of the plant $\mrm{G}(s)$ perturbed by disturbance $F\_{sd}$, is multiplied by cosine and sine of the disturbance frequency $\omega\_{r}$.
The cosine and sine multiplication of $\dot{x}$ results in
\begin{equation}\label{eq:demod_meas_signal_unfilt}
    \begin{bmatrix}
        \dot{x}\_{c}(t)\\
        \dot{x}\_{s}(t)
    \end{bmatrix}
    =
    \begin{bmatrix}
        2 \cos{(\omega\_{r}t + \psi\_{off})}\\
        2 \sin{(\omega\_{r}t + \psi\_{off})}
    \end{bmatrix}
    \dot{x}(t) \,,
\end{equation}
with $\dot{x}\_{c}$ and $\dot{x}\_{s}$ being the quadrature and in-phase components of the output and $\psi\_{off}$ as a phase offset.
It needs to be remarked that the factor of 2 in~\eqref{eq:demod_meas_signal_unfilt} follows the convention of~\cite{Lau2007}.

By making use of~\eqref{eq:laplace_cossin}-\eqref{eq:euler_formula}, the relation~\eqref{eq:demod_meas_signal_unfilt} results in the following frequency-domain representation
\begin{equation}\label{eq:demod_meas_signal_laplace}
    \begin{bmatrix}
        \mrm{\dot{X}}\_{c}(s)\\
        \mrm{\dot{X}}\_{s}(s)
    \end{bmatrix}
    =
    \left(
    e^{\rmj \psi\_{off}}
    \begin{bmatrix}
         1\\
        -\rmj 
    \end{bmatrix}
    \mrm{\dot{X}}(s_{-})
    +
    e^{-\rmj \psi\_{off}}
    \begin{bmatrix}
        1\\
        \rmj 
    \end{bmatrix}
    \mrm{\dot{X}}(s_{+}) 
    \right)
    \, .
\end{equation}
For each of the quadrature and in-phase channels, a linear time-invariant (LTI), SISO \textit{demodulated controller} $\mrm{C}(s)$ is implemented and forms a diagonal decoupled structure, as shown in the following
\begin{equation}\label{eq:control_diag}
    \begin{bmatrix}
        \mrm{\Delta T\_{g,c}}(s)\\
        \mrm{\Delta T\_{g,s}}(s)
    \end{bmatrix}
    =
    \mbf{C}(s)
    \begin{bmatrix}
        \mrm{\dot{X}}\_{c}(s)\\
        \mrm{\dot{X}}\_{s}(s)
    \end{bmatrix} \, ,
\end{equation}
where $\mbf{C}(s) = \mrm{C}(s) \mbf{I}_{2\times2}$.
The diagonal controller structure is intended for equal load reduction performance on both channels of the multivariable demodulated system and is valid if both channels have negligible interaction/coupling.
As will be shown in the following section, this condition is generally not met for all operating points. 
The phase offset earlier introduced in~\eqref{eq:demod_meas_signal_unfilt} plays an important role in the decoupling of the system throughout varying operational conditions.

Transforming back the quadrature and in-phase control signals $\mrm{\Delta T\_{g,c}}$ and $\mrm{\Delta T\_{g,s}}$ from the demodulated domain back to the additive torque signal $\mrm{\Delta T\_{g}}$ is accomplished by the modulation operation
\begin{equation}\label{eq:mod_freqdom}
    \begin{aligned}
        \mrm{\Delta T\_{g}}(s)
        &=
        \frac{1}{2}
        \left(
        \begin{bmatrix}
            1 & -\rmj
        \end{bmatrix}
        \begin{bmatrix}
            \mrm{\Delta T\_{g,c}}(s_{-}) \\
            \mrm{\Delta T\_{g,s}}(s_{-}) 
        \end{bmatrix}
        \right.\\
        &+
        \left.
        \begin{bmatrix}
            1 & \rmj
        \end{bmatrix}
        \begin{bmatrix}
            \mrm{\Delta T\_{g,c}}(s_{+}) \\
            \mrm{\Delta T\_{g,s}}(s_{+}) 
        \end{bmatrix}
        \right) \,,
    \end{aligned}
\end{equation}
which is the frequency domain equivalence of
\begin{equation}\label{eq:mod_timedom}
    \Delta T\_{g}(t)
    =
    \begin{bmatrix}
        \cos{(\omega\_{r}t)} & \sin{(\omega\_{r}t)}
    \end{bmatrix}
    \begin{bmatrix}
        \Delta T\_{g,c}(t)\\
        \Delta T\_{g,s}(t)
    \end{bmatrix}\, .
\end{equation}
The equation above completes the derivation for the MDC framework.

\begin{figure}[t!]
    \centering
    \subfloat[]{
        \includegraphics[width=0.975\linewidth,trim={0.1cm 0.1cm 0.1cm 0.1cm},clip]{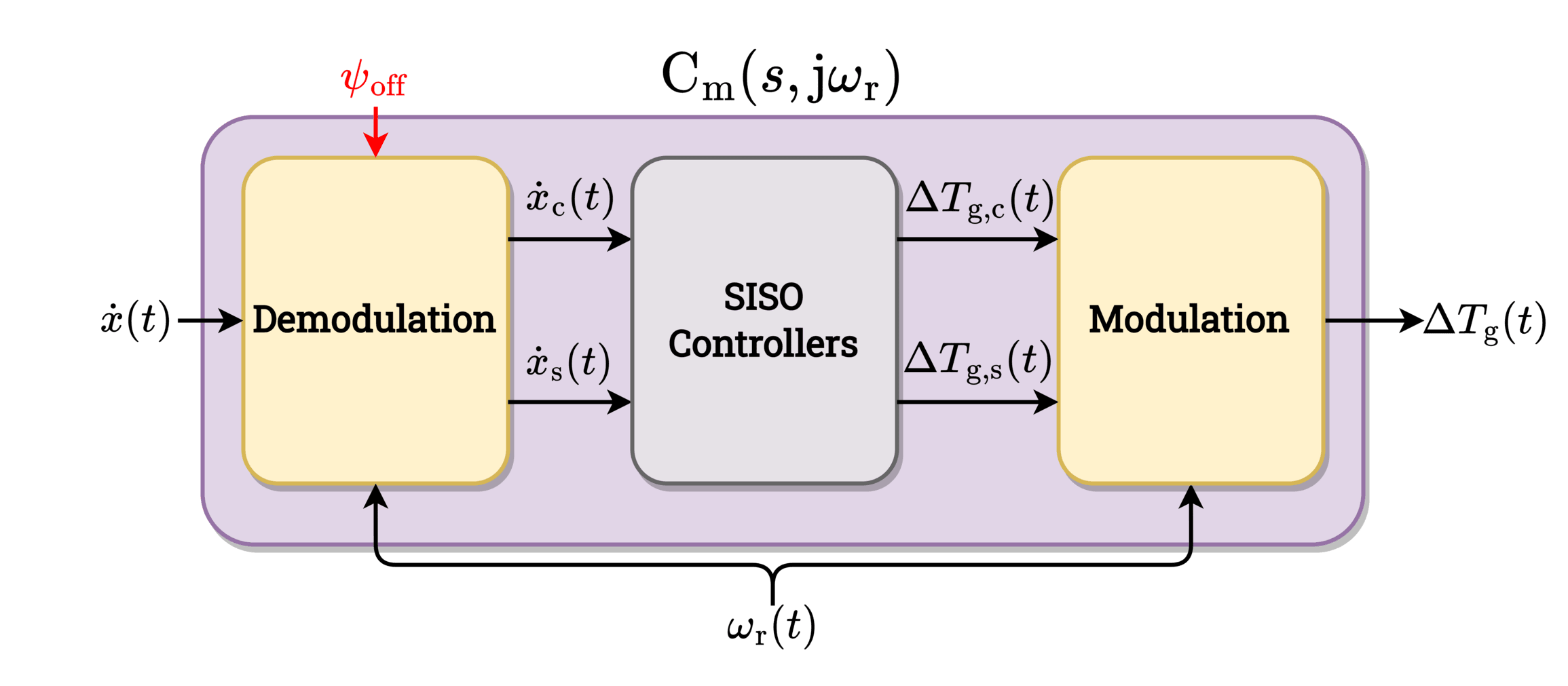}
        \label{fig:MDController}
    }
    \hfill
    \subfloat[]{
        \includegraphics[width=0.975\linewidth,trim={0.1cm 0.1cm 0.1cm 0.1cm},clip]{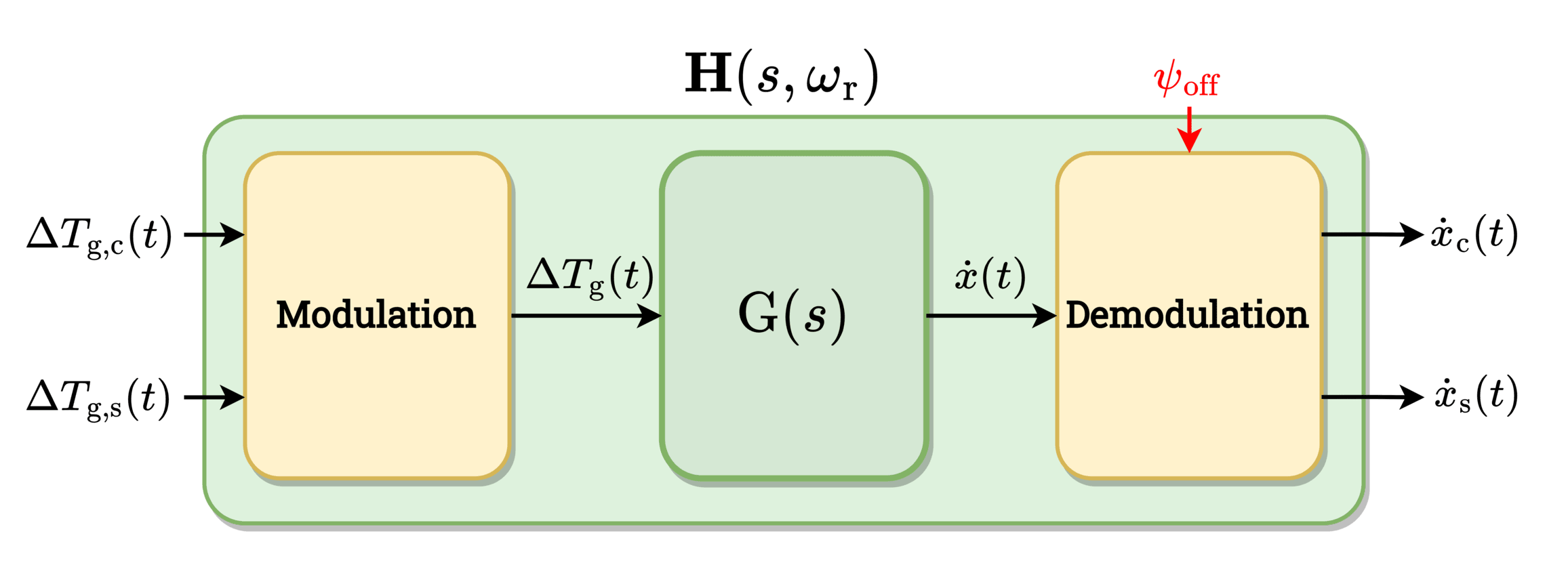}
         \label{fig:DemodPlant}
    }
    \caption{
    (a) SISO modulated controller $\mrm{C}\_{m}(s,\omega\_{r})$ and (b) MIMO demodulated plant $\mbf{H}(s,\omega\_{r})$ in the MDC scheme. 
    }
    \label{fig:ModulatedDemodulatedController}
\end{figure}

\section{MDC Controller and System Analysis}\label{sec:4}
\noindent
The established MDC framework allows for analysis of the system and controller, as depicted in Fig.~\ref{fig:ModulatedDemodulatedController}.
Figure~\ref{fig:MDController} shows the combination of the (de)modulators with the controller $\mbf{C}(s)$, forming a SISO \textit{modulated controller} representation from $\dot{x}$ to $\Delta T\_{g}$. 
This transformed controller possesses several beneficial properties, as further demonstrated in Section~\ref{sec:3_SISO_control}.
Presented in Fig.~\ref{fig:DemodPlant} is the multiple-input multiple-output (MIMO) \textit{demodulated plant} realization in the quadrature and in-phase channels from $[\Delta T\_{g,c},\Delta T\_{g,s}]^\top$ to $[\dot{x}\_{c},\dot{x}\_{s}]^\top$, resulting from the combination of $\mrm{G}(s)$ and the (de)modulators.
Of particular importance is the knowledge of potential cross-coupling between the demodulated channels of the MIMO plant presented in Section~\ref{sec:3_demod_plant}.
Section~\ref{sec:3_MDC4WT} provides insights into the properties of the demodulated multivariable system towards the justification of a decentralized controller $\mbf{C}(s)$.

The next subsections provide theoretical results for transformed controllers and systems, which are subsequently leveraged for the analysis of a linear wind turbine model. 
The phase offset $\psi\_{off}$ plays a remarkably important role in decoupling the demodulated system; however, to provide a clearer analysis, this variable will be included in the derivations after this section.

\subsection{SISO Modulated Controller Representation}\label{sec:3_SISO_control}
\noindent As previously indicated and shown in Fig.~\ref{fig:MDController}, the derived frequency-domain framework allows for a different perspective in analyzing the LTI controllers $\mbf{C}(s)$ in the modulation-demodulation scheme.
This section shows a remarkable property of the MDC scheme in that the LTI controllers are transformed into a SISO linear time-varying (LTV) controller structure when the modulation and demodulation stages are accounted for.

The SISO modulated controller representation from $\mrm{\dot{X}}(s)$ to $\mrm{\Delta T\_{g}}(s)$ is derived by first substituting~\eqref{eq:demod_meas_signal_laplace} to~\eqref{eq:control_diag} to obtain the following expression
\begin{equation}\label{eq:control_diag_subs}
    \begin{bmatrix}
        \mrm{\Delta T\_{g,c}}(s)\\
        \mrm{\Delta T\_{g,s}}(s)
    \end{bmatrix}
    =
    \mbf{C}(s)
    \left(
    \begin{bmatrix}
         1\\
        -\rmj 
    \end{bmatrix}
    \mrm{\dot{X}}(s_{-})
    +
    \begin{bmatrix}
        1\\
        \rmj 
    \end{bmatrix}
    \mrm{\dot{X}}(s_{+}) 
    \right)
    \,,
\end{equation}
which is subsequently combined with~\eqref{eq:mod_freqdom}, resulting in
\begin{equation}\label{eq:mod_freqdom_subs}
    \mrm{\Delta T\_{g}}(s) = \mrm{C}\_{m}(s,\omega\_{r}) \mrm{\dot{X}}(s) = (\mrm{C}(s_{-}) + \mrm{C}(s_{+})) \mrm{\dot{X}}(s)\,,
\end{equation}
being scheduled by $\omega\_{r}$.
The above relation in~\eqref{eq:mod_freqdom_subs} shows that simple LTI controllers in the demodulated system become LTV if the (de)modulators are included.
Under the assumption that $\omega\_{r}$ is slowly varying and does not (significantly) change within its period, as stated previously, the results from the linear analysis framework in this section generalize to the nonlinear implementation.

Using the derived relation between $\mrm{C}(s)$ and $\mrm{C}\_{m}(s,\omega\_{r})$ in~\eqref{eq:mod_freqdom_subs}.
This section shows three, $n = \{1,2,3\}$, controller types of interest $\mrm{C}_{n}(s)$, for which a convenient analytical expression $\mrm{C}_{\mrm{m},n}(s,\omega\_{r})$ exists:
\begin{enumerate}
    \item{
    The first LTI controller is a \textit{proportional controller}
    \begin{equation}\label{eq:prop_control}
        \mrm{C}_{1} = K\_{P} \,,
    \end{equation}
    with $K\_{P}\in\mathbb{R}$ as a constant gain, produces $\Delta T\_{g,c}$ and $\Delta T\_{g,s}$ that scale $\dot{x}\_{c}$ and $\dot{x}\_{s}$.
    This controller is transformed into
    \begin{equation}\label{eq:prop_control_Cm}
        \mrm{C}_{\mrm{m},1} = 2 K\_{P} \,,
    \end{equation}
    independent from $\omega\_{r}$---thus, retains the LTI characteristic of $\mrm{C}_{1}$.
    }
    
    \item{
    The second LTI controller is an \textit{integral controller}
    \begin{equation}\label{eq:int_control}
        \mrm{C}_{2}(s) = \frac{K\_{I}}{s} \,,
    \end{equation}
    with $K\_{I}\in\mathbb{R}$ as an integral gain, which has infinite gain for steady-state deviations and alleviates high-frequent components of $\dot{x}\_{c}$ and $\dot{x}\_{s}$.
    Transformation of the considered controller results in
    \begin{equation}\label{eq:int_control_Cm}
        \mrm{C}_{\mrm{m},2}(s,\omega\_{r}) = \frac{2 K\_{I} s}{s^2+\omega\_{r}^2} \,,
    \end{equation}
    being an undamped inverted notch filter with a complex pole pair at $\pm \rmj \omega\_{r}$ and, thus, infinite gain at $\omega\_{r}$~\cite{Bodson1994}.
    This also means that full cancellation of periodic load at this frequency is possible by this type of controller.
    Also, it is worth mentioning that this controller structure bears a similarity with that of repetitive control.
    However, both controllers differ in that~\eqref{eq:int_control_Cm} only tackles the fundamental disturbance frequency, whereas repetitive control also inherently accounts for the higher harmonics.
    The interested reader is referred to the literature, e.g.~\cite{Steinbuch2002,Liu2022b}, for more details on repetitive control.
    }
    
    \item{
    The last LTI controller is a \textit{first-order low-pass filter}
    \begin{equation}\label{eq:LPF_first}
        \mrm{C}_{3}(s) = \frac{K\_{L}}{s+\omega\_{LPF}} \,,
    \end{equation}
    with a constant gain $K\_{L}\in\mathbb{R}$ and $\omega\_{LPF}\in\mathbb{R}^+$ as the cut-off frequency.
    This controller is similar to the previous two LTI controllers in that it proportionally scales $\dot{x}\_{c}$ and $\dot{x}\_{s}$ but also alleviates high-frequent components in the demodulated measurement signals.
    The introduction of $\omega\_{LPF}$ moves the pole away from the origin, resulting in a non-infinite steady-state gain, in contrast to the integrator controller.
    Its SISO modulated expression
    \begin{multline}\label{eq:LPF_first_Cm}
        \mrm{C}_{\mrm{m},3}(s,\omega\_{r}) = 
        \frac{2 K\_{L} (s + \omega\_{LPF})}
        {s^2 + 2 \omega\_{LPF}s + \omega\_{LPF}^2 + \omega\_{r}^2} \,,
    \end{multline}
    interestingly, is also an inverted notch but with a beneficial property in that it contains a damping term in the denominator ($2\omega\_{LPF}s$), tunable via the selection of $\omega\_{LPF}$.
    This means that, in contrast to $\mrm{C}_{\mrm{m},2}(s,\omega\_{r})$, the gain of the controller at $\omega\_{r}$ can be limited, which can be desirable in terms of actuation activity needed to dampen periodic loading and added robustness.
    }
\end{enumerate}

\begin{figure}[!t]
    \centering
    \includegraphics[width=0.95\linewidth]{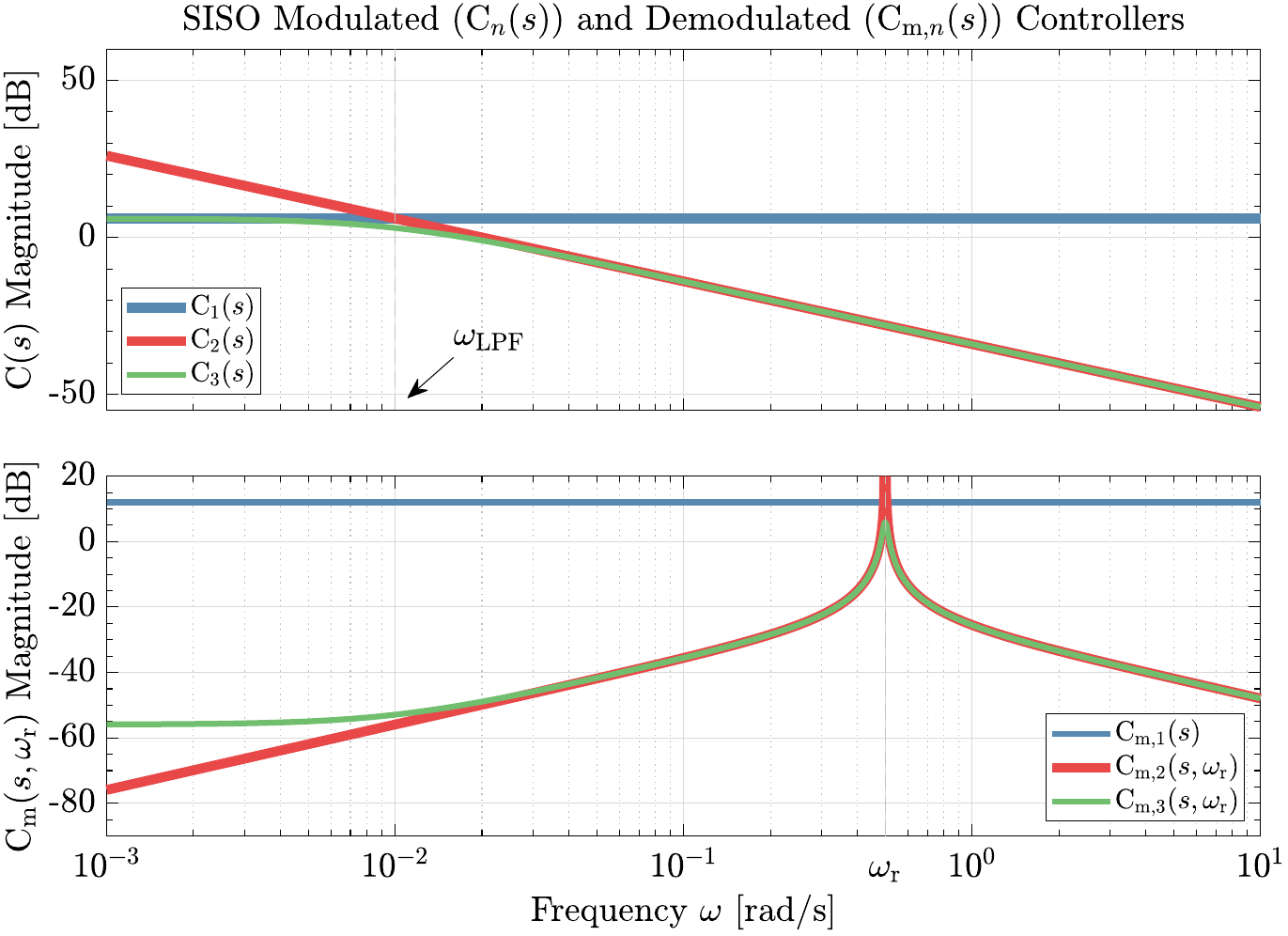}
    \caption{
    Bode magnitude plots of the demodulated controller $\mrm{C}_{n}(s)$ (top) and the resulting SISO modulated controller $\mrm{C}_{\mrm{m},n}(s, \omega\_{r})$ (bottom), $n = \{1,2,3\}$.
    The proportional controller $\mrm{C}_{1}$ is mapped into $\mrm{C}_{\mrm{m},1}$, being the same proportional controller but with an additional factor of 2.
    The integral controller $\mrm{C}_{2}(s)$ is rendered into an undamped inverted notch filter $\mrm{C}_{\mrm{m},2}(s, \omega\_{r})$ with infinite gain at $\omega\_{r} = 0.5$~rad/s.
    The low-pass filter $\mrm{C}_{3}(s)$ with a cut-off frequency of $\omega\_{LPF} = 0.01$~rad/s results into a damped inverted notch filter $\mrm{C}_{\mrm{m},3}(s, \omega\_{r})$.
    }
    \label{fig:bodePlotControllers}
\end{figure}

Figure~\ref{fig:bodePlotControllers} depicts the Bode magnitude plots of the considered nominal and transformed controllers, which are, respectively, shown by the top and bottom plots to support the discussed observed conclusions.
The plots are created with arbitrary choices for $\omega\_{r} = 0.5$~rad/s,  $\omega\_{LPF} = 0.01$~rad/s and ${K\_{P} = 2}$.
For the gains in $\mrm{C}_{2}(s)$ and $\mrm{C}_{3}(s)$, ${K\_{I} = K\_{L} = 2 \omega\_{LPF}}$ is chosen such that its crossover frequency matches with that of $\mrm{C}_{3}(s)$, which results in a clearer comparison.

In addition to the aforementioned presented LTI controllers $\mrm{C}_{n}(s)$, alternative controller configurations, potentially of greater complexity, could be suggested and their SISO modulated (LTV) counterparts could be derived by using~\eqref{eq:mod_freqdom_subs}.

\subsection{MIMO Demodulated Plant Representation}\label{sec:3_demod_plant}
\noindent
The incorporation of the (de)modulators in the proposed framework enables the plant dynamics to be represented as a MIMO system, as shown in Fig.~\ref{fig:DemodPlant}, with which the presence of cross-coupling between channels can be investigated.
To render the plant dynamics into their demodulated, MIMO representation, the SISO plant $\mrm{G}(s)$ is substituted into~\eqref{eq:demod_meas_signal_laplace} such that the following equation is obtained
\begin{equation}\label{eq:MIMO_plant1}
    \begin{gathered}
        \begin{bmatrix}
            \mrm{\dot{X}}\_{c}(s)\\
            \mrm{\dot{X}}\_{s}(s)
        \end{bmatrix}
        =
        \begin{bmatrix}
             1\\
            -\rmj
        \end{bmatrix}
        \mrm{G}(s_{-}) \mrm{\Delta T}\_{g}(s_{-})
        +
        \begin{bmatrix}
            1\\
            \rmj
        \end{bmatrix}
        \mrm{G}(s_{+}) \mrm{\Delta T}\_{g}(s_{+})
        \, ,
    \end{gathered}
\end{equation}
and substituting~\eqref{eq:mod_freqdom} into~\eqref{eq:MIMO_plant1} gives rise to the 2P (i.e. twice-per-revolution) terms indicated by $s_{2\pm} = s \pm 2\rmj \omega\_{r}$ as follows
\begin{equation}\label{eq:Tg2ddotx_Coleman}
    \begin{gathered}
        \begin{bmatrix}
            \mrm{\dot{X}}\_{c}(s)\\
            \mrm{\dot{X}}\_{s}(s)
        \end{bmatrix}
        =
        \underbrace{
        \begin{bmatrix}
            \mbf{G}_1(s,\omega\_{r})^\top \\ \mbf{G}_2(s,\omega\_{r})^\top \\ \mbf{G}_3(s,\omega\_{r})^\top
        \end{bmatrix}^\top}_{\mbf{H}(s,\omega\_{r})^\top}
        \begin{bmatrix}
            \mrm{\Delta T\_{g,c}}(s_{2-}) \\
            \mrm{\Delta T\_{g,s}}(s_{2-}) \\
            \mrm{\Delta T\_{g,c}}(s) \\
            \mrm{\Delta T\_{g,s}}(s) \\
            \mrm{\Delta T\_{g,c}}(s_{2+}) \\
            \mrm{\Delta T\_{g,s}}(s_{2+}) 
        \end{bmatrix}
        \,,
    \end{gathered}
\end{equation}
with $\mbf{H}(s,\omega\_{r})$ as the concatenation of the following transfer matrices 
\begin{subequations}\label{eq:G123s}
\begin{align}
    \mbf{G}_1(s,\omega\_{r}) &=
    \frac{1}{2}
    \begin{bmatrix}
        \mrm{G}(s_{-})           & -\rmj \mrm{G}(s_{-})\\
        -\rmj \mrm{G}(s_{-}) & -\mrm{G}(s_{-})
    \end{bmatrix}\,,\\
    \mbf{G}_2(s,\omega\_{r}) &=
    \frac{1}{2}
    \begin{bmatrix}
        \mrm{G}(s_{-})        + \mrm{G}(s_{+})  & \rmj \mrm{G}(s_{-})  - \mrm{G}(s_{+})\\
        \rmj (-\mrm{G}(s_{-}) + \mrm{G}(s_{+})) & \mrm{G}(s_{-})        + \mrm{G}(s_{+})          
    \end{bmatrix}\,,\label{eq:G2s}\\
    \mbf{G}_3(s,\omega\_{r}) &=
    \frac{1}{2}
    \begin{bmatrix}
        \mrm{G}(s_{+})      & \rmj \mrm{G}(s_{+})\\
        \rmj \mrm{G}(s_{+}) & -\mrm{G}(s_{+})
    \end{bmatrix}\,.
\end{align}
\end{subequations}

The higher harmonic terms add complexity to the control design and analysis since all the contributions $\mbf{G}_1(s,\omega\_{r})$, $\mbf{G}_2(s,\omega\_{r})$, and $\mbf{G}_3(s,\omega\_{r})$ need to be accounted.
Therefore, the relation above is simplified by selecting an appropriate controller structure that filters out 2P frequency components such that several terms can be omitted, i.e., $[\Delta T\_{g,c}(s_{2\pm}),\Delta T\_{g,s}(s_{2\pm})]^\top\approx 0$.
Among the LTI controllers presented in~Section~\ref{sec:3_SISO_control}, either $\mrm{C}_{2}(s)$ or $\mrm{C}_{3}(s)$ is a viable candidate due to the roll-off at high frequencies.
Therefore~\eqref{eq:Tg2ddotx_Coleman} simplifies into
\begin{equation}\label{eq:mod_demod_WT}
    \begin{bmatrix}
        \mrm{\dot{X}}\_{c}(s)\\
        \mrm{\dot{X}}\_{s}(s)
    \end{bmatrix}
    \approx
    \mbf{G}_2(s,\omega\_{r}) 
    \begin{bmatrix}
        \mrm{\Delta T\_{g,c}}(s) \\
        \mrm{\Delta T\_{g,s}}(s) 
    \end{bmatrix}
    \,,
\end{equation}
representing an approximation of the demodulated multivariable plant.

\subsection{Application of MDC on a Simplified Wind Turbine}\label{sec:3_MDC4WT}
\noindent 
Now, with the tower dynamics transfer function $\mrm{G}(s)$ at hand, the definition of $\mbf{G}_{2}(s,\omega\_{r})$ in~\eqref{eq:G123s} is used to transform the nominal dynamics into its demodulated counterpart. 
It is compelling to investigate their relations by studying Bode plots of both their dynamics. 
To this end, rather arbitrary wind turbine modal parameters ${m = 3\times10^4}$~kg, ${d\_{eff} = 3\times10^3}$~Ns/m, and ${k = 1.5\times10^4}$~N/m are considered.
The chosen parameters resemble a soft-soft wind turbine tower with its natural frequency being ${\omega\_{n}=\sqrt{k/m}=0.7071}$~rad/s and the rotor operates in $\omega\_{r} \in \Omega = [\omega\_{r,min},\omega\_{r,rated}]$, with ${\omega\_{r,min}=0.5}$~rad/s and ${\omega\_{r,rated}=1.2}$~rad/s.

\begin{figure*}[!t]
    \centering
    \subfloat[][]{
        \includegraphics[width=0.48\linewidth]{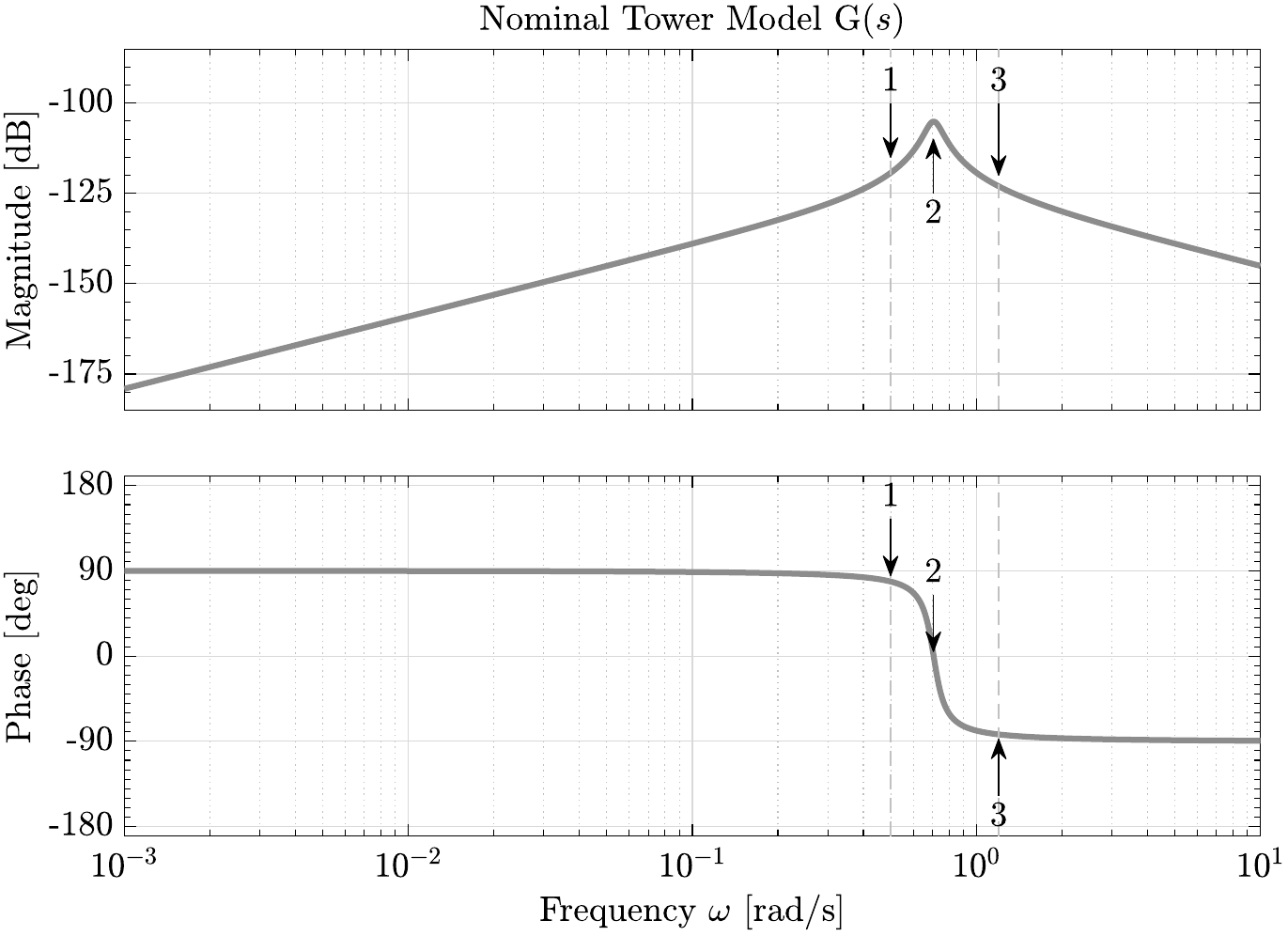}
        \label{fig:bode_Gs}
    }
    \subfloat[]{
        \includegraphics[width=0.489\linewidth]{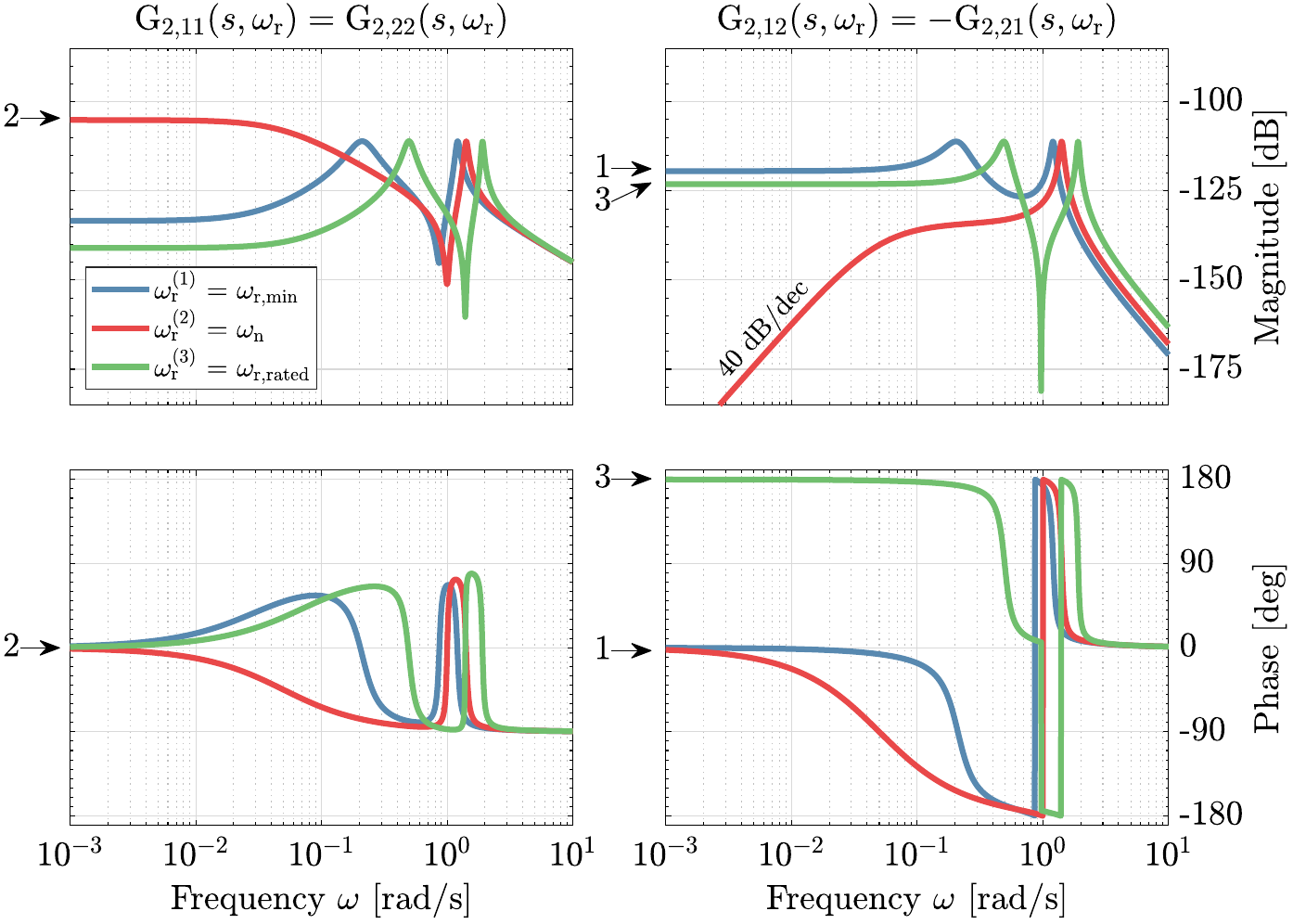}
         \label{fig:bode_G2s}
    }
    \caption{
    Bode magnitude and phase plots of the nominal~(a) and the demodulated~(b) wind turbine models.
    In~(a), vertical dashed lines indicate the operating range of a soft-soft wind turbine $\mrm{G}(s)$, where a resonance peak at about $\omega = \omega\_{n} = 0.7071$~rad/s is apparent in the magnitude plot.
    A $180^\circ$ phase shift occurs due to the presence of this resonance, as shown in the corresponding phase plot.
    The points indicated by labeled arrows $i=\{1,2,3\}$ represent three sample points $\omega\_{r}^{(i)} = \{\omega\_{r,min}, \omega\_{n}, \omega\_{r,rated}\}$, with ${\omega\_{r,min} = 0.5}$~rad/s and ${\omega\_{r,rated} = 1.2}$~rad/s, to evaluate the mapping from $\mrm{G}(s)$ into the steady-state components of $\mbf{G}_{2}(s,\omega\_{r})$, as shown in~(b), before, during, and after the resonance.
    Note the $40$~dB/dec slope in the magnitude plot of $\mrm{G}_{2,12}(s,\omega\_{r}^{(2)})$ at low frequencies, which indicates the presence of two zeros at the origin.
    }
    \label{fig:bodePlots}
\end{figure*}

Figure~\ref{fig:bodePlots} shows the Bode plots of both $\mrm{G}(s)$ and $\mbf{G}_{2}(s,\omega\_{r})$, in which the former and latter transfer functions are represented respectively in Figs.~\ref{fig:bode_Gs} and~\ref{fig:bode_G2s}.
As $\mbf{G}_{2}(s,\omega\_{r})$ is a $2\times2$ (skew-symmetric) transfer function matrix, its main and off-diagonal elements, 
\begin{equation}\label{eq:G2md}
    \begin{aligned}
        \mrm{G}_{2,11}(s,\omega\_{r})&=\mrm{G}_{2,22}(s,\omega\_{r}) \\
        &=\frac{1}{2} \frac{\mrm{N}(s\_{-})\mrm{D}(s\_{+})+\mrm{N}(s\_{+})\mrm{D}(s\_{-})}{\mrm{D}(s\_{-})\mrm{D}(s\_{+})} \,,
    \end{aligned}
\end{equation}
and 
\begin{equation}\label{eq:G2od}
    \begin{aligned}
        \mrm{G}_{2,12}(s,\omega\_{r})&=-\mrm{G}_{2,21}(s,\omega\_{r}) \\
        &= \frac{\rmj}{2} \frac{\mrm{N}(s\_{-})\mrm{D}(s\_{+})-\mrm{N}(s\_{+})\mrm{D}(s\_{-})}{\mrm{D}(s\_{-})\mrm{D}(s\_{+})} \,,
    \end{aligned}
\end{equation}
are shown in the respective first and second columns of Fig.~\ref{fig:bode_G2s}.
Since $\mbf{G}_{2}(s,\omega\_{r})$ is parameterized by the 1P frequency, its Bode plot is evaluated for different $\omega\_{r}$ values.
Specifically, in this case, three frequencies are considered to understand the system's behavior before, at, and after the tower's resonance frequency, i.e., $\omega\_{r}^{(i)} = \{\omega\_{r,min}, \omega\_{n}, \omega\_{r,rated}\}$, where $i=\{1,2,3\}$.

The modulation-demodulation of $\mrm{G}(s)$ at $s = \rmj \omega\_{r}^{(i)}$ maps its magnitude, $|\mrm{G}(\rmj \omega\_{r}^{(i)})|$, into the steady-state magnitude of the demodulated plant, $|\mbf{G}_{2}(0,\omega\_{r}^{(i)})|$.
Roughly speaking, for the diagonal LTI SISO controller $\mbf{C}(s)$ to be justified for the entire turbine operating range, it is desirable to have the steady-state magnitudes of the main diagonal be more dominant than the off-diagonal counterparts, that is 
\begin{equation}\label{eq:md_dominance}
    |\mrm{G}_{2,11}(0,\omega\_{r})| \gg |\mrm{G}_{2,12}(0,\omega\_{r})| \,, \quad \forall \omega\_{r} \in \Omega \,.
\end{equation}
When the above condition is met, the quadrature and in-phase channels are well decoupled and thus, no significant interaction at low-frequency region and steady-state between channels is present~\cite{Lau2007}.

In Fig.~\ref{fig:bode_G2s}, it is shown that this is the case for $\mbf{G}_{2}(0,\omega\_{r}^{(2)})$, with $\omega\_{r}^{(2)} = \omega\_{n}$.
The reason behind such main-diagonal dominance at this operating point is partly due to the presence of a pair of zeros (differentiators) at the origin of $\mrm{G}_{2,12}(s,\omega\_{r}^{(2)})$.
Consequently, $|\mrm{G}_{2,12}(s,\omega\_{r}^{(2)})|$ is $0$ at steady-state and increases with a $40$~dB/dec slope as frequency goes higher, also indicated in the figure.
Moreover, as $|\mrm{G}_{2,11}(s,\omega\_{r}^{(2)})|$ has a maximized steady-state contribution, shown by the flat-line region at low frequencies, it can be directly concluded that~\eqref{eq:md_dominance} is satisfied in this case.
This is not necessarily the case when $\omega\_{r}\neq\omega\_{n}$, as exemplified by $\omega\_{r}^{(1)}$ and $\omega\_{r}^{(3)}$.
In both cases, the absence of any differential and integral actions in $\mrm{G}_{2,12}(s,\omega\_{r})$ results in flat magnitudes at low frequencies but with higher gains compared to $\mrm{G}_{2,11}(s,\omega\_{r})$; thus, cross-coupling is present for the lower frequency region of interest.
Hence, it may be preferred to utilize $\mrm{G}_{2,12}(s,\omega\_{r})$ for control due to these higher gains at these operating conditions.
Nevertheless, additional complexity arises in doing so as a $180^\circ$ of phase difference presents in $\angle \mrm{G}_{2,12}(0,\omega\_{r})$ (shown by arrows 1 and 3 in the figure).
This infers that a gain sign flip occurs when the turbine switches operating regime from $\omega\_{r}<\omega\_{n}$ to $\omega\_{r}>\omega\_{n}$ and vice versa, which may result in instability.

Also noticeable in Fig.~\ref{fig:bode_G2s} is the presence of two resonance peaks in $|\mbf{G}_{2}(s,\omega\_{r})|$ for $\omega\_{r}^{(1)}$ and $\omega\_{r}^{(3)}$ with their magnitudes being $-6$~dB lower than that of $|\mrm{G}(\rmj \omega\_{n})|$.
These two peaks originate from the shift of the nominal plant's natural frequency into $|\omega\_{n}\pm\omega\_{r}^{(i)}|$ due to the $\mrm{D}(s_\pm)$ terms in~\eqref{eq:G2md}-\eqref{eq:G2od}, which also explains why only a single peak presents for $\omega\_{r}^{(2)}$ case~\cite{Lau2007}.
Although there is significant coupling in higher frequencies in the off-diagonal, only the low-frequent region is of interest for the controller design.

The above observation on the magnitude and phase mapping between both plant representations can be understood better by taking another look at~\eqref{eq:G2md}-\eqref{eq:G2od}.
First, since both $\mrm{G}_{2,11}(s,\omega\_{r})$ and $\mrm{G}_{2,12}(s,\omega\_{r})$ are constituted by the same poles, whether or not~\eqref{eq:md_dominance} is satisfied depends only on the zeros of these transfer functions.
Secondly, as these zeros are located at $z_{1,2}=\pm\sqrt{\omega\_{n}-\omega\_{r}}$, they can be either purely on the imaginary axis, origin, or real axis, depending on the value of the rotational speed of the turbine with respect to the tower's natural frequency.
This creates different (steady-state) phase behavior for $\omega\_{r}<\omega\_{n}$ and $\omega\_{r}>\omega\_{n}$ because the latter produces a right-half plane (RHP) zero at the dominant channel such that the aforementioned $180^\circ$ phase difference/sign flip occurs.
Therefore, to ensure main-diagonal dominance for the entire turbine operating range and elimination of the phase drop in the dominant channels, it becomes compelling to manipulate the zero locations of $\mrm{G}_{2,11}(s,\omega\_{r})$ and $\mrm{G}_{2,12}(s,\omega\_{r})$.
In the following section, both goals can be achieved simultaneously by the inclusion of an offset in the MDC scheme.

\begin{rem}\label{rem:different_measurements}
    Whether $\ddot{x}$, $\dot{x}$, or $x$ is used as the output of the plant affects the numerator of $\mrm{G}(s)$ and thus the zeros of~\eqref{eq:G2md}-\eqref{eq:G2od}.
    This also determines how~\eqref{eq:md_dominance} is satisfied for different operating points.
    Nevertheless, regardless of the selected output signal, channel cross-coupling and a $180^\circ$ phase shift in the dominant channel of $\mbf{G}_{2}(s,\omega\_{r})$ still exists, which necessitates their compensation by phase offset inclusion.
\end{rem}

\section{Quadrature and In-Phase Channels Decoupling by Phase Offset Inclusion}\label{sec:5} 
\noindent 
The MIMO demodulated plant channel cross-coupling, as well as the gain sign flip examined earlier, has uncovered potential challenges in the proposed MDC design.
To gain more knowledge on the degree of this coupling for the entire turbine operating range, a more reliable metric, namely the relative gain array, is used in Section~\ref{sec:4_RGA}.
As inferred in the previous section, the inclusion of the phase offset $\psi\_{off}$ in the MDC plays a key role in the decoupling of the MIMO demodulated plant.
In Section~\ref{sec:4_phase_offset}, the optimal phase offset value, by which the highest degree of decoupling and gain sign flip correction can be achieved, is discussed.

\subsection{Relative Gain Array Analysis}\label{sec:4_RGA}
\noindent 
Relative gain array (RGA), denoted $\mbf{\Lambda}(\cdot)$, is a measure of interaction between multiple control channels~\cite{Skogestad2005}.
The RGA is used to assess the coupling of the MIMO demodulated system at steady-state $\mbf{G}_{2}(0,\omega\_{r})$ as follows:
\begin{equation}\label{eq:RGA_G}
    \mbf{\Lambda}(\mbf{G}_{2}(0,\omega\_{r})) = \mbf{G}_{2}(0,\omega\_{r}) \circ \mbf{G}_{2}(0,\omega\_{r})^{-\top} \,,
\end{equation}
where `$\circ$' denotes an element-by-element multiplication known as the Hadamard or Schur product.

\begin{figure*}[!t]
     \centering
    \subfloat[]{
        \label{fig:RGAsteadystate}
        \includegraphics[width=0.48\linewidth]{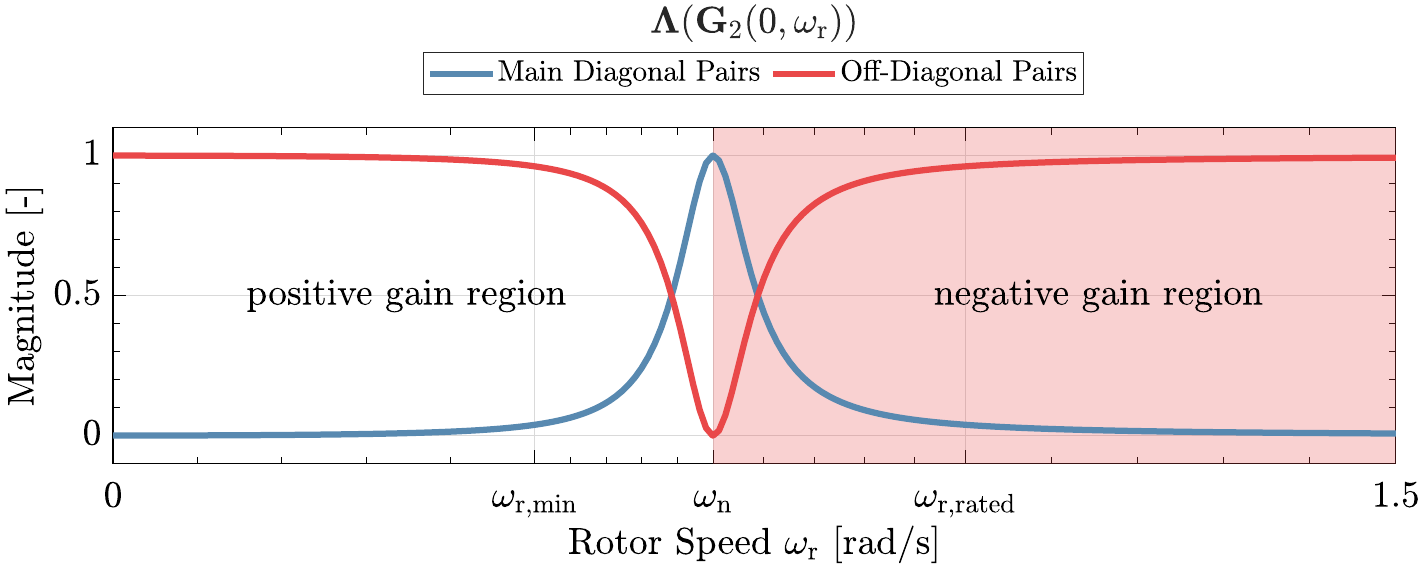}
    }
     \hfill
    \subfloat[]{
    \label{fig:RGAsteadystateOffset}
        \includegraphics[width=0.48\linewidth]{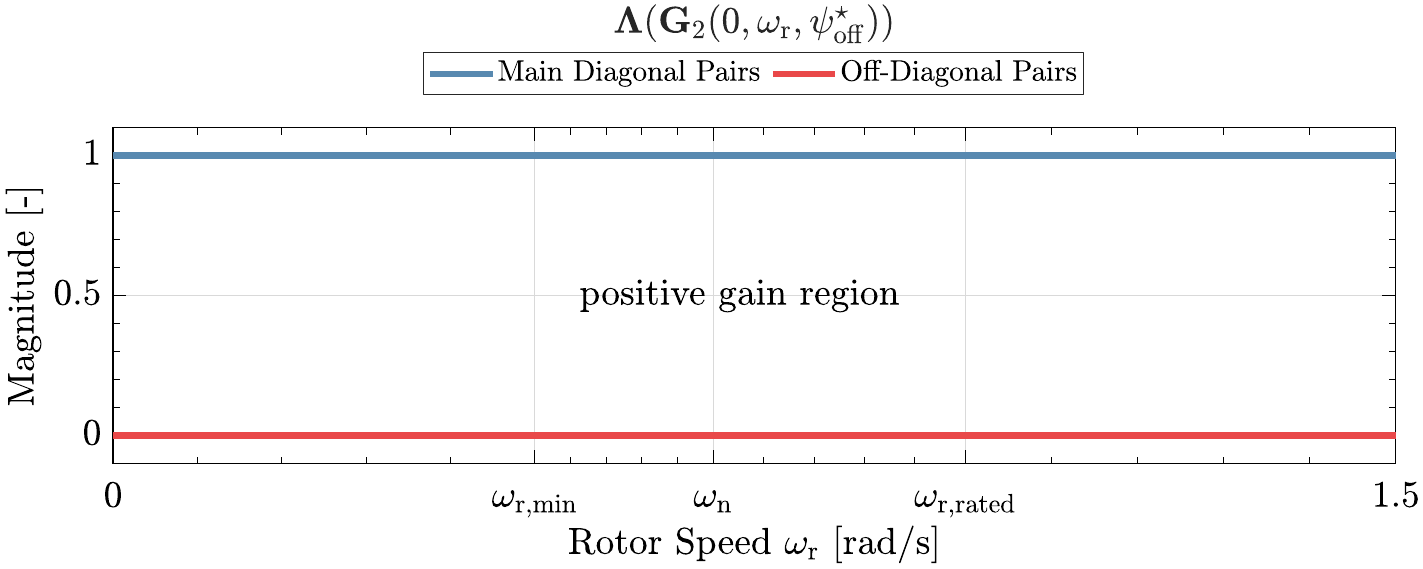}
    }        
    \caption{
    Steady-state RGA of $\mbf{G}_{2}(s,\omega\_{r})$ over the operating regime of a simple wind turbine model without~(a) and with~(b) phase offset $\psi\_{off}$.
    The inclusion of optimal phase offset $\psi\_{off}^\star$ into the MDC results in decoupled quadrature and in-phase input-output channels at steady-state, as well as eliminating the need for control gain swapping.
    Please note that for the figures, $\dot{x}$ is measured.
    In case of $x$ or $\ddot{x}$, these results, in particular that of~(a) will be different (see Remark~\ref{rem:different_measurements}).
    }
    \label{fig:RGAs}
\end{figure*}

Figure~\ref{fig:RGAs} shows the evaluation of $\mbf{\Lambda}(\mbf{G}_{2}(0,\omega\_{r}))$ for an extended range of rotor operation {$\omega\_{r} \in \Omega' = [0,1.5]$~rad/s}, where the magnitude of the main diagonal elements ${|\Lambda_{11}| = |\Lambda_{22}|}$ is shown by the blue lines and that of the off-diagonal ${|\Lambda_{12}| = |\Lambda_{21}|}$ is represented by the red lines.
As the rows and columns of $\mbf{\Lambda}(\mbf{G}_{2}(0,\omega\_{r}))$ sum to 1, it is sufficient to mention only $|\Lambda_{11}|$ for the following discussion.

Figure~\ref{fig:RGAsteadystate} depicts the current case where the main diagonal pairings are dominant with ${|\Lambda_{11}| = 1}$ only about $\omega\_{r}^{(2)}=\omega\_{n}$.
Also evident is the increasing off-diagonal dominance as $\omega\_{r}$ deviates from $\omega\_{n}$ with ${|\Lambda\_{11}| \approx 0}$ at $\omega\_{r}^{(1)}$ and $\omega\_{r}^{(3)}$.
This shows agreement with the previous Bode plot observations in Fig.~\ref{fig:bode_G2s} and hints that the current input-output pairings preference is not suitable for the entire operating range~\cite{Skogestad2005}.
Swapping the input-output pairings to the off-diagonal may be preferable but insufficient to account for the negative gain resulting from the $180^\circ$ phase difference in $\angle \mbf{G}_{2}(0,\omega\_{r})$ indicated by the red-shaded region.
Figure~\ref{fig:RGAsteadystateOffset} shows an ideal case where ${|\Lambda_{11}| = 1}$ for the entire operating regime without any gain sign change, as opposed to Fig.~\ref{fig:RGAsteadystate}.
In the following section, such a condition is shown to be achievable by means of phase offset inclusion in the proposed MDC framework.

\subsection{Phase Offset Inclusion}\label{sec:4_phase_offset}
\noindent
In Section~\ref{sec:3_MDC4WT}, it has been shown that the value of $\omega\_{r}$ plays a role in the positioning of the zeros of $\mrm{G}_{2,11}(s,\omega\_{r})$ and $\mrm{G}_{2,12}(s,\omega\_{r})$, resulting in both channel cross-coupling and gain sign flip.
The previously omitted phase offset $\psi\_{off}$, however, may play a critical role in tackling both issues at the same time by influencing these zero locations.
In particular, the optimal phase offset value, defined by
\begin{equation}\label{eq:opt_phase_offset}
    \psi\_{off}^\star(\omega\_{r}) = -\angle \mrm{G}(\rmj \omega\_{r}) \,,
\end{equation}
can be chosen.
As the plant's dynamics vary according to the frequency of the periodic excitation, $\psi\_{off}^\star$ varies according to $\omega\_{r}$.
In the remainder of this paper, the notation $\omega\_{r}$ is dropped when referring to $\psi\_{off}^\star$ for brevity's sake.
This offset value has been rigorously studied in the literature, where methods such as averaging theory, root locus, and loop-shaping have been employed.
The interested reader is referred to~\cite{Byl2005,Messner1995}, and references therein for more detailed analysis.
The effects of $\psi\_{off}$ inclusion on the MIMO demodulated plant and SISO modulated controller are discussed in Sections~\ref{sec:4_phase_off_plant} and \ref{sec:4_phase_offset_ctrl}.

\subsubsection{Effects of $\psi\_{off}$ on MIMO Demodulated Plant}\label{sec:4_phase_off_plant}
\noindent
To understand the effects $\psi\_{off}$ creates on $\mbf{G}_2(s,\omega\_{r})$, the derivation done in Section~\ref{sec:3_demod_plant} is repeated by including this offset, which results in the following relation
\begin{multline}\label{eq:G2s_offset}
    \mbf{G}_2(s,\omega\_{r},\psi\_{off}^\star) =\\
    \begin{bmatrix}
        \frac{e^{\rmj \psi\_{off}^\star} \mrm{G}(s_{-})        + e^{-\rmj \psi\_{off}^\star} \mrm{G}(s_{+}) }{2} & \rmj \frac{e^{\rmj \psi\_{off}^\star} \mrm{G}(s_{-})  - e^{-\rmj \psi\_{off}^\star} \mrm{G}(s_{+})}{2}\\
        \rmj \frac{(-e^{\rmj \psi\_{off}^\star} \mrm{G}(s_{-}) + e^{-\rmj \psi\_{off}^\star} \mrm{G}(s_{+}))}{2} & \frac{e^{\rmj \psi\_{off}^\star} \mrm{G}(s_{-})        + e^{-\rmj \psi\_{off}^\star} \mrm{G}(s_{+}) }{2}
    \end{bmatrix}\,,
\end{multline}
in which $\psi\_{off}=\psi\_{off}^\star$ is applied.
In comparison with the original definition of $\mbf{G}_{2}(s,\omega\_{r})$ in~\eqref{eq:G2s}, $e^{\pm\rmj\psi\_{off}^\star}$ terms appear in~\eqref{eq:G2s_offset} after the inclusion of $\psi\_{off}^\star$ into the MDC scheme.
These terms play a role in zero positioning of both $\mrm{G}_{2,11}(s,\omega\_{r},\psi\_{off}^\star)$ and $\mrm{G}_{2,12}(s,\omega\_{r},\psi\_{off}^\star)$, thereby affecting their (steady-state) gains and phases.

\begin{figure}[t!]
    \centering
    \includegraphics[width=\linewidth]{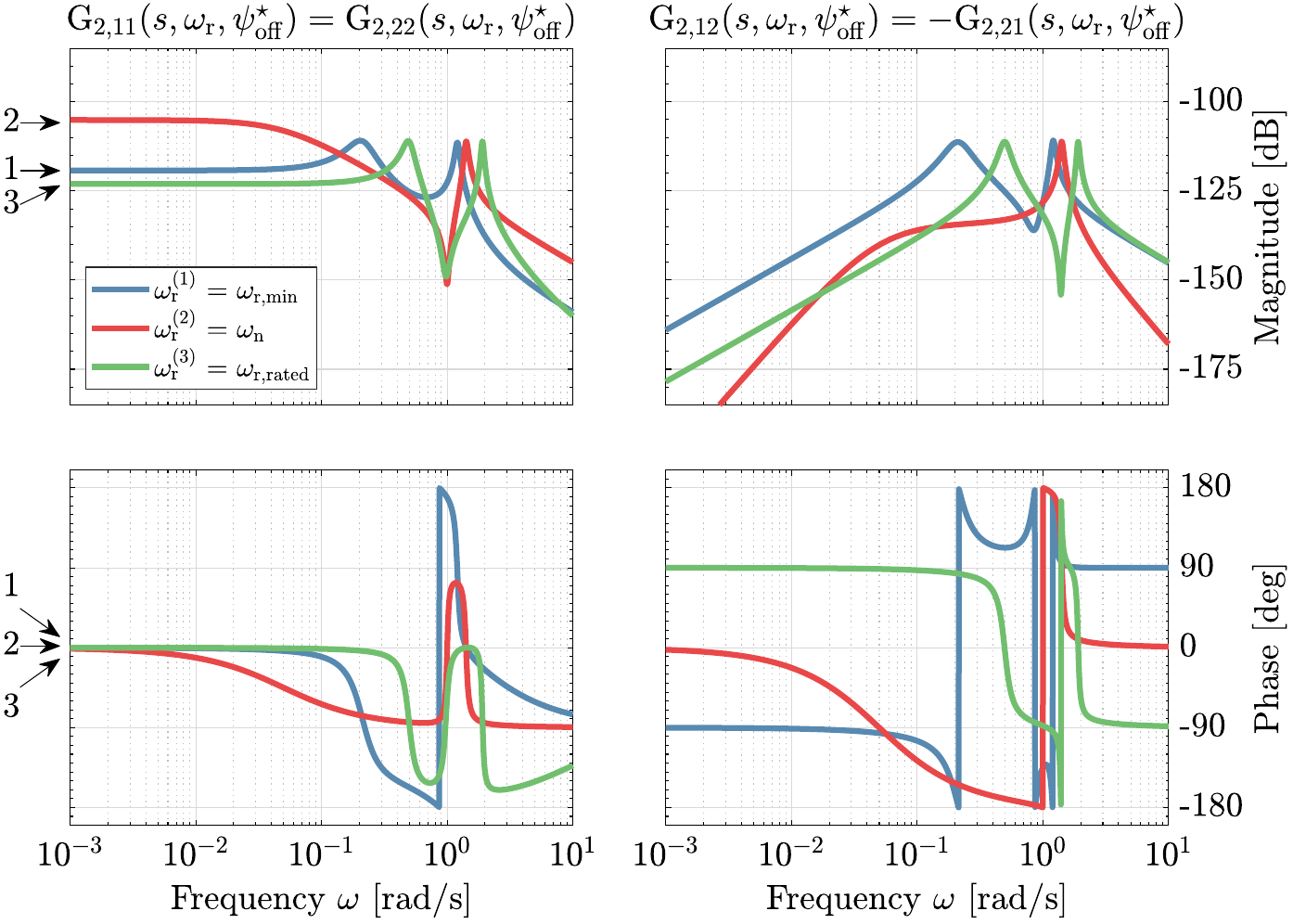}
    \caption{
    Bode plot of the demodulated wind turbine model with the optimal phase offset included $\mbf{G}_{2}(s,\omega\_{r},\psi\_{off}^\star)$.
    It is evident that the steady-state off-diagonal contributions are attenuated and that the main diagonal components become dominant.
    }
    \label{fig:bode_G2s_offset}
\end{figure}

Figure~\ref{fig:bode_G2s_offset} depicts the Bode plot of the MIMO demodulated wind turbine model including $\psi\_{off}^\star$~\eqref{eq:G2s_offset}.
It is apparent that in comparison with the previous case in Fig.~\ref{fig:bode_G2s}, the transfer function matrix has now become diagonally dominant, with their phases starting from zero at the steady state and not exhibiting $180^\circ$ phase difference anymore.
This main-diagonal dominance is made clearer by investigating the analytical expressions for the main and off-diagonal elements of the MIMO demodulated plant at steady-state by substituting~\eqref{eq:Gs} into~\eqref{eq:G2s_offset} and setting $s = 0$~rad/s, that is
\begin{multline}\label{eq_app:symbolic_G2s_md_0offset_ss}
    \mrm{G}_{2,11}(0,\omega\_{r},\psi\_{off}^\star) = \\
    s\_{f} 
    \frac{m \sin{(\psi\_{off}^\star)} \omega\_{r}^3+d \cos{(\psi\_{off}^\star)} \omega\_{r}^2-k \sin{(\psi\_{off}^\star)} \omega\_{r}}{d^2 \omega\_{r}^2+k^2-2 k m \omega\_{r}^2+m^2 \omega\_{r}^4} \,,
\end{multline}
and
\begin{multline}\label{eq_app:symbolic_G2s_od_0offset_ss}
    \mrm{G}_{2,12}(0,\omega\_{r},\psi\_{off}^\star) = \\
    s\_{f} 
    \frac{-m \cos{(\psi\_{off}^\star)} \omega\_{r}^3+d \sin{(\psi\_{off}^\star)} \omega\_{r}^2+k \cos{(\psi\_{off}^\star)} \omega\_{r}}{d^2 \omega\_{r}^2+k^2-2 k m \omega\_{r}^2+m^2 \omega\_{r}^4} \,,
\end{multline}
respectively.
Then, the steady-state magnitudes of both the main and off-diagonal elements can be computed for all operating points, where  the main diagonal's magnitude equals that of the nominal plant at the excitation frequency 
\begin{equation}
    |\mrm{G}_{2,11}(0,\omega\_{r},\psi\_{off}^\star)| = |\mrm{G}(\rmj \omega\_{r})| \,,
\end{equation}
whereas
\begin{equation}
    |\mrm{G}\_{2,12}(0,\omega\_{r},\psi\_{off}^\star)| = 0 \,.
\end{equation}
This means that~\eqref{eq:md_dominance} is always fulfilled.

Steady-state RGA evaluation of the MIMO demodulated plant after the optimal offset inclusion $\mbf{\Lambda}(\mbf{G}(0,\omega\_{r},\psi\_{off}^\star))$ also confirms the above observation.
This is depicted in Fig.~\ref{fig:RGAsteadystateOffset}, where ${|\Lambda_{11}| = 1}$ for the entire (extended) operating range.

The above observations conclude that under the inclusion of $\psi\_{off}^\star$, the main diagonal dominance is asserted and no gain sign flip is experienced as the rotational frequency sweeps through the tower's natural frequency.
Therefore, the use of diagonal controller $\mbf{C}(s)$ is now justified. 

\subsubsection{Effects of $\psi\_{off}$ on SISO Modulated Controller}\label{sec:4_phase_offset_ctrl}
Similar to the previous section, the effects of the $\psi\_{off}$ inclusion on the SISO modulated controller requires $\mrm{C}\_{m}(s,\omega\_{r})$ derived in Section~\ref{sec:3_SISO_control} to be reformulated into
\begin{equation}\label{eq:mod_freqdom_subs_offset}
    \mrm{C}\_{m}(s,\omega\_{r},\psi\_{off}) = 
    e^{-\rmj \psi\_{off}}
    \mrm{C}(s_{-})
    +
    e^{\rmj \psi\_{off}}
    \mrm{C}(s_{+})
    \,.
\end{equation}
Effectively, the LTI controllers~\eqref{eq:prop_control}, \eqref{eq:int_control}, and~\eqref{eq:LPF_first} previously proposed are transformed  by~\eqref{eq:mod_freqdom_subs_offset} (also by making use of~\eqref{eq:euler_formula}) into the following respective (LTV) controllers:
\begin{equation}\label{eq:prop_control_Cm_offset}
    \mrm{C}_{\mrm{m},1}(\psi\_{off}) = 2 K\_{P} \cos{(\psi\_{off})} \,, 
\end{equation}
\begin{equation}\label{eq:int_control_Cm_offset}
    \mrm{C}_{\mrm{m},2}(s,\omega\_{r},\psi\_{off}) = \frac{2 K\_{I} (\cos{(\psi\_{off})} s + \sin{(\psi\_{off})} \omega\_{r})}{s^2+\omega\_{r}^2} \,,
\end{equation}
\begin{multline}\label{eq:LPF_first_Cm_offset}
    \mrm{C}_{\mrm{m},3}(s,\omega\_{r},\psi\_{off}) = \\ 
    \frac{2 K\_{L} ( \cos{(\psi\_{off})} (s + \omega\_{LPF}) + \sin{(\psi\_{off})} \omega\_{r} )}
    {s^2 + 2 \omega\_{LPF} s + \omega\_{LPF}^2 + \omega\_{r}^2} \,.
\end{multline}

From~\eqref{eq:prop_control_Cm_offset}-\eqref{eq:LPF_first_Cm_offset}, it can be seen that the phase offset $\psi\_{off}$ influences the modulated controllers in the following ways.
First, for~$\mrm{C}_{\mrm{m},1}(\psi\_{off})$, the phase offset affects the gain of the controller.
However, in Section~\ref{sec:3_demod_plant}, it has been stated that this controller structure cannot filter out the 2P frequency components; therefore, it is not considered any further during the time-domain demonstration in the next section. 
For~$\mrm{C}_{\mrm{m},2}(s,\omega\_{r},\psi\_{off})$, its zero location becomes
\begin{equation*}
    z_{\mrm{m},2} = -\omega\_{r} \tan{(\psi\_{off})} \,.
\end{equation*}
In the original formulation~\eqref{eq:int_control_Cm}, $\mrm{C}_{\mrm{m},2}(s,\omega\_{r})$ has a pure zero at the origin but the offset enables relocation of this zero into the left- or right-half plane (LHP or RHP).
Similarly, for~$\mrm{C}_{\mrm{m},3}(s,\omega\_{r},\psi\_{off})$, its zero is relocated from $-\omega\_{LPF}$ into
\begin{equation*}
    z_{\mrm{m},3} = -\omega\_{LPF}-\omega\_{r} \tan{(\psi\_{off})} \,.
\end{equation*}
An in-depth analysis of the controller zero positioning by this offset is discussed in~\cite{Byl2005}.

\begin{figure}[!t]
    \centering
    \includegraphics[width=0.95\linewidth]{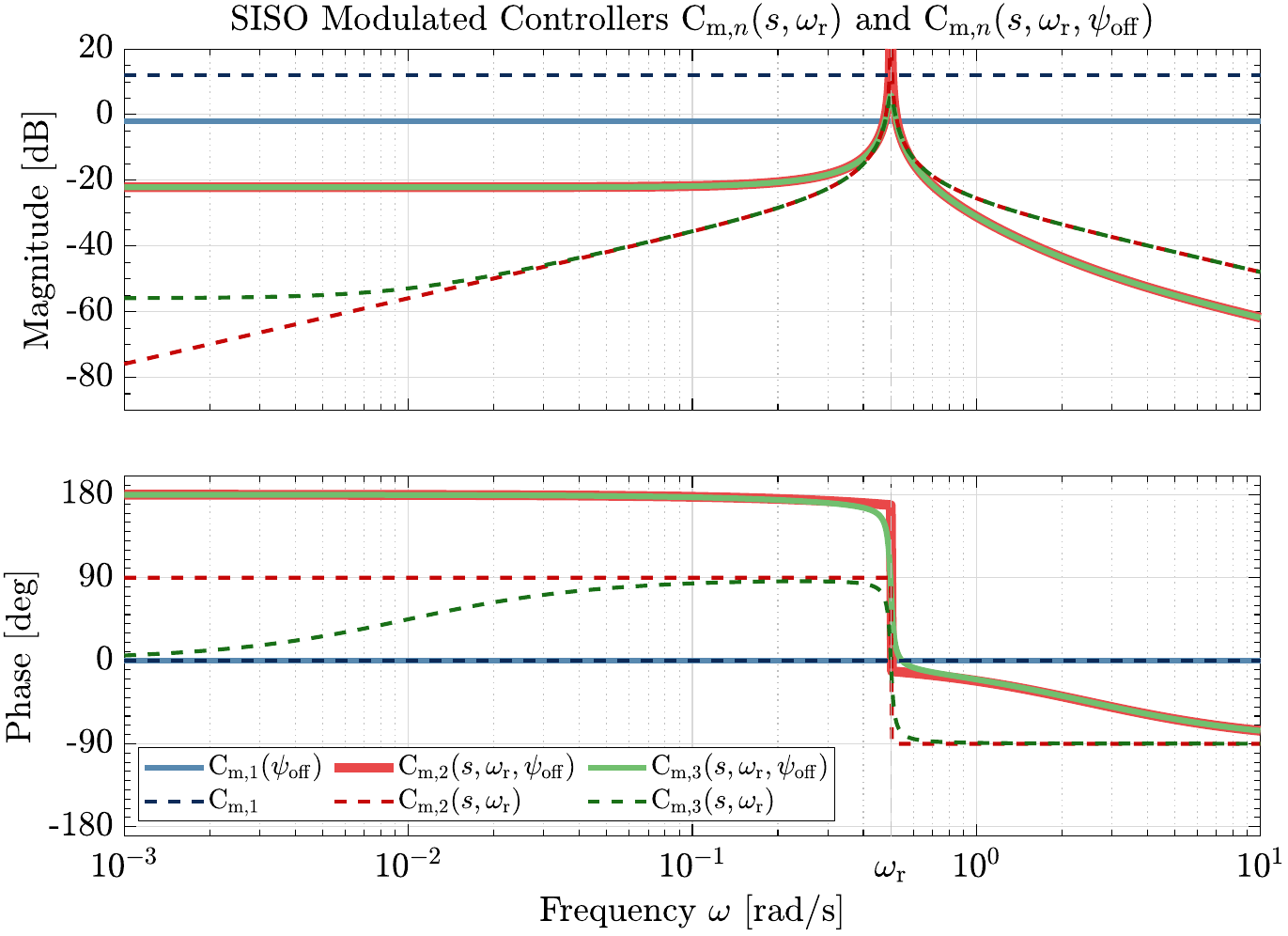}
    \caption{
    Bode plots of SISO modulated controllers, without (dashed lines) and with (solid lines) $\psi\_{off}$, $\mrm{C}_{\mrm{m},n}(s, \omega\_{r})$ and $\mrm{C}_{\mrm{m},n}(s, \omega\_{r}, \psi\_{off})$, respectively, $n=\{1,2,3\}$ and ${\omega\_{r} = 0.5}$~rad/s.
    }
    \label{fig:bodePlotControllersOffset}
\end{figure}

Figure~\ref{fig:bodePlotControllersOffset} illustrates the Bode plots of the SISO modulated controllers, similar to that of Section~\ref{sec:3_SISO_control}, without and with $\psi\_{off}$.
Compared to the inverted notch filters without $\psi\_{off}$, those with the optimal offset included exhibit increased magnitude at the low frequencies due to the introduction of a zero (for $\mrm{C}_{\mrm{m},2}(s,\omega\_{r},\psi\_{off})$) or relocation of an existing zero to a high frequency (for $\mrm{C}_{\mrm{m},3}(s,\omega\_{r},\psi\_{off})$).
For both controllers, the same $\psi\_{off}$ is chosen.

\section{Simulation Results}\label{sec:6} 
\noindent
In this section, simulations demonstrating the performance of the proposed control scheme are carried out.
The control scheme consists of the conventional active tower damping controller for increasing the effective damping of the side-side tower motion, as explained in Section~\ref{sec:2_baseline_ctrl}. 
This conventional controller is augmented by the MDC studied earlier in Sections~\ref{sec:4} and~\ref{sec:5} to alleviate the periodic 1P fatigue load.

Time-domain simulations at two fidelity levels are considered, where the lower fidelity simulations, discussed in Section~\ref{sec:6_simple_turbine}, show the proof-of-concept of the MDC framework with the simplified wind turbine model derived earlier.
Afterward, high-fidelity simulations employing the National Renewable Energy Laboratory (NREL) OpenFAST software package~\cite{Jonkman2021} are covered in Section~\ref{sec:6_OpenFAST_turbine}.

\subsection{Simplified Turbine Simulations}\label{sec:6_simple_turbine}
\noindent
For the simulations presented here, the wind turbine model derived in Section~\ref{sec:2} is employed.
The synthetic tower properties used in Section~\ref{sec:3_MDC4WT} are utilized for the tower dynamics~\eqref{eq:dynamics_tower}.
The parameters of NREL 5-MW reference wind turbine~\cite{Jonkman2009}, as shown in Table~\ref{tab:NREL-5MW}, are used for the rotor dynamics~\eqref{eq:dynamics_drivetrain} and to determine the below-rated torque controller gain according to~\eqref{eq:Komega2}.

A steady, uniform, staircase wind inflow from ${v = 5}$~m/s to ${v = 10}$~m/s with $1.25$~m/s of speed increment is generated for the simulations.
Each wind speed lasts for $250$~s, resulting in $1250$~s of total simulation time.
The choice of this wind speed condition is made such that the rotor starts about $\omega\_{r,min}$ at ${\omega\_{r} = 0.58}$~rad/s and ends near $\omega\_{r,rated}$ at ${\omega\_{r} = 1.128}$~rad/s, thus covering most of the operating range $\Omega$ and that resonance is experienced when $v = 6.25$~m/s at $t = 250-500$~s.
To model a rotor imbalance, $F\_{sd}$ with $a\_{sd} = 150$~N and $\phi\_{sd} = \pi/4$~rad is selected, equivalent to $a\_{sd}/s\_{f} = 9$~kNm of torque amplitude at the tower-top.

\begin{table}[t!]
    \caption{Parameters of the (modified) NREL-5MW reference wind turbine and environment condition.}
    \centering
    \begin{tabular}{llrl} 
        \hline
        \textbf{Description} & \textbf{Notation} & \textbf{Value} & \textbf{Unit} \\ [0.5ex] 
        \hline\hline
        Rated generator power               & -                   & 5                       & MW        \\     
        Optimal tip-speed ratio             & $\lambda^\star$     & 7                       & -             \\
        Max. power coefficient              & $C\_{p}^\star$      & 0.458                   & -             \\
        Fine pitch angle                    & -                   & 0                       & $^\circ$      \\
        Optimal torque gain (LSS)           & $K$                 & 2.1286 $\cdot 10^6$     & Nm/(rad/s)$^{2}$      \\
        LSS equivalent inertia              & $J\_{r}$            & 4.0802$\cdot 10^7$      & kgm$^2$ \\ 
        Gearbox ratio                       & $G$                 & 97                      & - \\ 
        Rotor radius                        & $R$                 & 63                      & m  \\
        Minimum rotor speed                 & $\omega\_{r,min}$   & 0.5                     & rad/s       \\             
        Rated rotor speed                   & $\omega\_{r,rated}$ & 1.2                     & rad/s       \\             
        Rated generator torque              & $T\_{g,rated}$      & 43.09355                & kNm        \\   
        Generator efficiency                & -                   & 0.944                   & -             \\
        Hub height                          & $H$                 & 90                      & m  \\
        Tow. natural frequency (sca.)       & $\omega\_{n,s}$       & 0.6963                  & rad/s\\ 
        Tow. modal mass (sca.)              & $m\_{s}$            & 3.6200$\cdot 10^5$      & kg  \\
        Tow. modal damping (sca.)           & $d\_{s}$            & 2.4588$\cdot 10^3$     & Ns/m \\
        Tow. modal stiffness (sca.)         & $k\_{s}$            & 1.7677$\cdot 10^5$      & N/m \\
        Air density                         & $\rho$              & 1.225                   & kg/m$^3$  \\ 
        [1ex] 
        \hline
        syn. = synthetic; sca. = scaled
    \end{tabular}
    \label{tab:NREL-5MW}
\end{table}

\begin{figure}[!t]
    \centering
    \includegraphics[width=\linewidth]{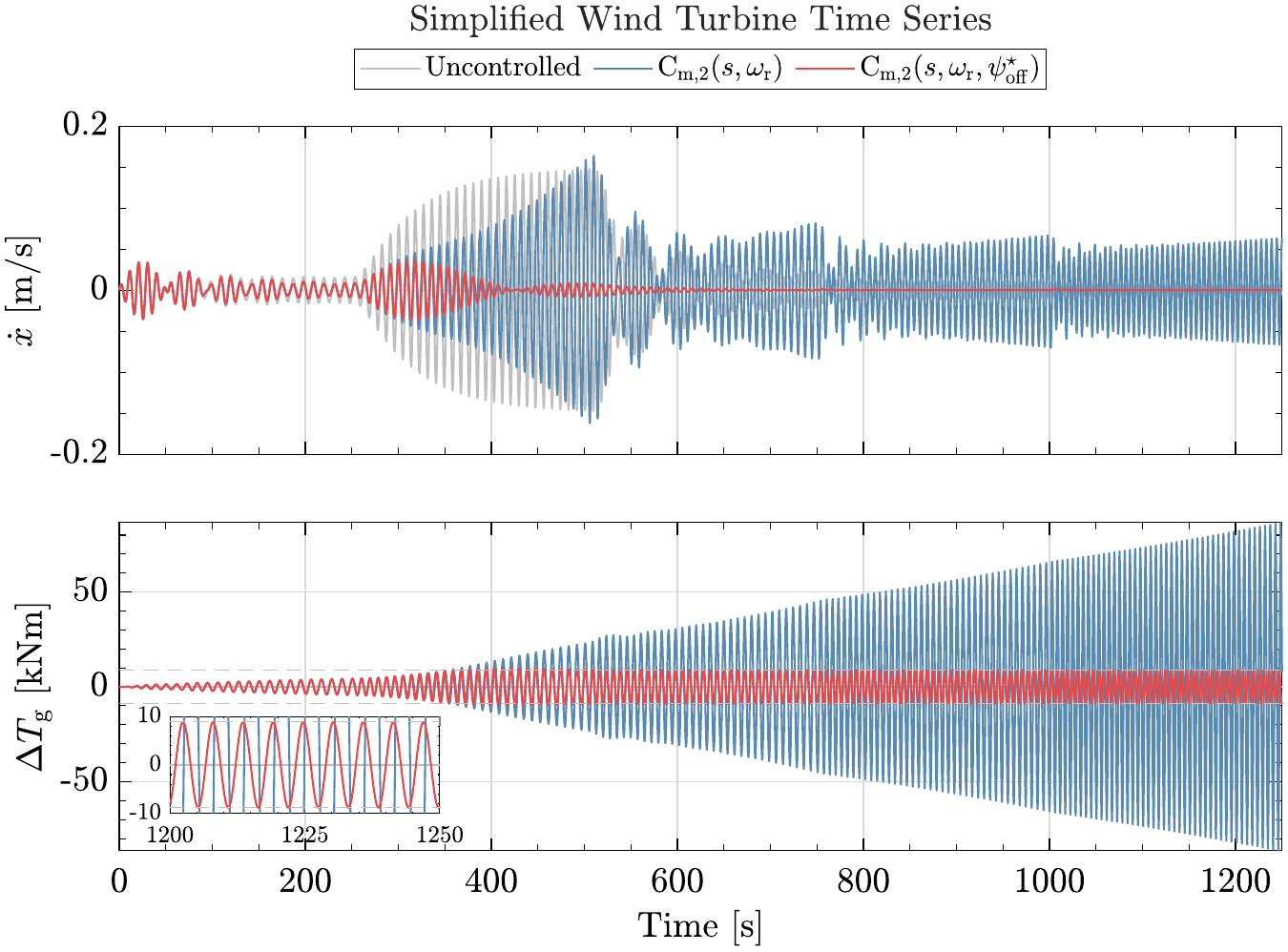}
    \caption{
    Time series response of tower velocity (top) and additive generator torque (bottom) under staircase wind ${v = 5-10}$~m/s where the performance of $\mrm{C}_{\mrm{m},2}(s,\omega\_{r})$ and $\mrm{C}_{\mrm{m},2}(s,\omega\_{r},\psi\_{off}^\star)$ are demonstrated.
    Horizontal dashed lines of $\pm9$~kNm in the bottom plot indicate the periodic load magnitude's equivalence in terms of torque.
    A zoomed-in plot depicts control action behavior at the end of the simulation.
    }
    \label{fig:time_series_simple_Cm2}
\end{figure}

\begin{figure}[!t]
    \centering
    \includegraphics[width=\linewidth]{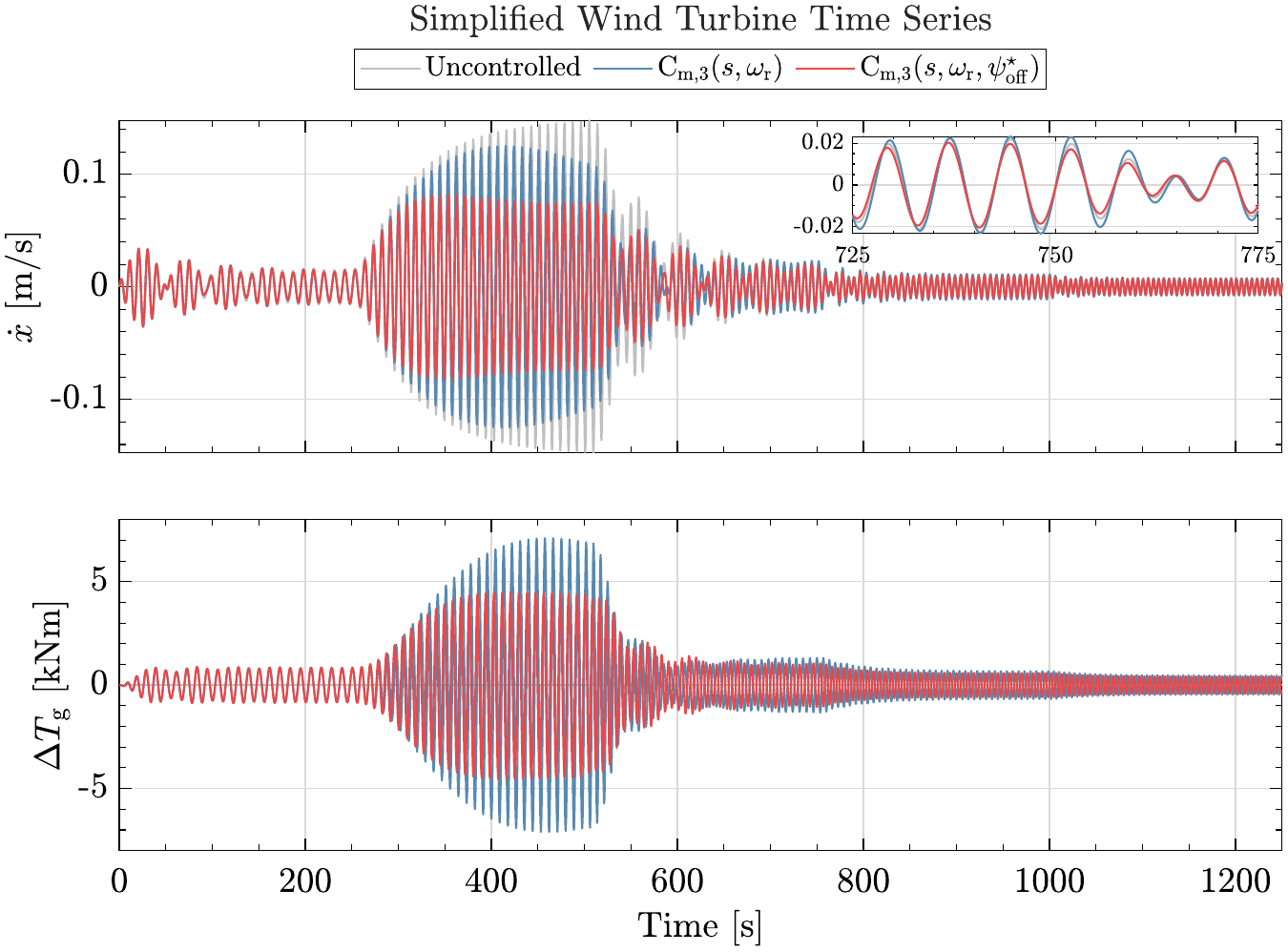}
    \caption{
    Time series response of tower velocity (top) and additive generator torque (bottom) under staircase wind ${v = 5-10}$~m/s where the performance of $\mrm{C}_{\mrm{m},3}(s,\omega\_{r})$ and $\mrm{C}_{\mrm{m},3}(s,\omega\_{r},\psi\_{off}^\star)$ are demonstrated.
    A zoomed-in plot depicts tower velocities at $t = 725-750$~s.
    }
    \label{fig:time_series_simple_Cm3}
\end{figure}

Figures~\ref{fig:time_series_simple_Cm2}-\ref{fig:time_series_simple_Cm3} depict the performance of the controllers in the MDC scheme, where only the (damped) inverted notch filters are of interest, without and with $\psi\_{off}^\star$ included.
The respective blue and red lines show the former and latter MDCs, whereas the gray lines show the uncontrolled wind turbine responses.
In the figures, $\dot{x}$ and $\Delta T\_{g}$ measurements are depicted by the top and bottom plots, respectively.
For MDCs without $\psi\_{off}^\star$, ${\psi\_{off}=-90^\circ}$ is used for the whole operating range $\Omega$ to swap the input-output pairings to the more dominant off-diagonal pairs (see Section~\ref{sec:4_RGA}).
For brevity's sake, the notation $\mrm{C}_{\mrm{m},n}(s,\omega\_{r})$ is kept for referring to the pair-swapped MDCs in this section.

The tuning gains for the controllers are chosen to be ${K\_{I} = K\_{L} = 1500}$ and ${\omega\_{LPF} = 0.025}$~rad/s while ensuring stability.
During the simulations, the value of $\psi\_{off}^\star$ is determined by means of a look-up table (LUT), fed by filtered rotor measurements where a first-order LPF with a cut-off frequency of $0.2$~rad/s is employed. 

Figure~\ref{fig:time_series_simple_Cm2} compares the performance of the undamped inverted notch filters $\mrm{C}_{\mrm{m},2}(s,\omega\_{r})$ and $\mrm{C}_{\mrm{m},2}(s,\omega\_{r},\psi\_{off}^\star)$.
It is observed from the figure that $\mrm{C}_{\mrm{m},2}(s,\omega\_{r})$ does not cancel the 1P periodic loading at the tower as shown in the measurements of $\dot{x}$.
During resonance, tower oscillation starts to grow due to the strong coupling at this frequency as the main-diagonal pairings gain dominance (see Fig.~\ref{fig:RGAsteadystate}).
After the resonance, the controller enters the negative gain region (i.e., sign flip occurs) and the growth of $\Delta T\_{g}$ becomes unbounded.
On the other hand, $\mrm{C}_{\mrm{m},2}(s,\omega\_{r},\psi\_{off}^\star)$ does not exhibit instability and fully cancels the 1P load.
The full cancellation of the periodic load is attributed to the infinite gain of the controller at the disturbance frequency.
Notice the convergence of the control action's amplitude to $9$~kNm (equal to $a\_{sd}/s\_{f}$) as indicated by the horizontal dashed lines and a zoomed-in plot for ${t=1200-1250}$~s.

Figure~\ref{fig:time_series_simple_Cm3} depicts the performance of the damped inverted notch filters $\mrm{C}_{\mrm{m},3}(s,\omega\_{r})$ and $\mrm{C}_{\mrm{m},3}(s,\omega\_{r},\psi\_{off}^\star)$.
The differences in both controllers' performance are evident once the wind speed reaches ${v = 6.25}$~m/s, where resonance starts to occur.
The former is shown to dampen the tower's oscillation at about $t = 400 - 450$~s (and slightly beyond when ${v = 7.5}$~m/s is reached), however not as effective as the latter, shown by the greater reduction in $\dot{x}$ with lower control action.
An inset plot at the top highlights that after the resonance, exemplified for $t = 725-750$~s, a slight increase in tower oscillation is caused by $\mrm{C}_{\mrm{m},3}(s,\omega\_{r})$.
On the other hand, evident tower motion reduction is performed by $\mrm{C}_{\mrm{m},3}(s,\omega\_{r},\psi\_{off}^\star)$.
In comparison to $\mrm{C}_{\mrm{m},2}(s,\omega\_{r},\psi\_{off}^\star)$ in Fig.~\ref{fig:time_series_simple_Cm2}, $\mrm{C}_{\mrm{m},3}(s,\omega\_{r},\psi\_{off}^\star)$ does not fully cancel the 1P periodic loading due to limited gain at $\omega\_{r}$.
Nonetheless, maximum $\Delta T\_{g}$ magnitude of only $\approx4.5$~kNm is observed in $\mrm{C}_{\mrm{m},3}(s,\omega\_{r},\psi\_{off}^\star)$ contrast to aggressive $9$~kNm exhibited by $\mrm{C}_{\mrm{m},2}(s,\omega\_{r},\psi\_{off}^\star)$.

From these simplified wind turbine simulations, it is concluded that $\psi\_{off}^\star$ is crucial in the load-mitigating performance of the proposed MDCs and in preventing closed-loop instability.
Secondly, it can be observed that MDCs perform best in terms of 1P load reduction when $\omega\_{r}=\omega\_{n}$ as their gains are highest at this frequency.
This motivates a gain-scheduling strategy to be incorporated into the framework, done in the higher fidelity simulations of the following section.

\subsection{OpenFAST Simulations}\label{sec:6_OpenFAST_turbine}
\noindent
In the high-fidelity OpenFAST simulations presented in this section, the NREL 5-MW reference wind turbine is again used.
However, since this reference turbine's tower was originally classified as soft-stiff, its wall thickness is downscaled by a factor of $7.5$ to recast it into a soft-soft tower design.
This consequently reduces tower mass so that its first natural frequency approximates ${\omega\_{n} = 0.7071}$~rad/s of the soft-soft tower in the simplified wind turbine setting.
The tower modal mass, damping, and stiffness for this scaled tower are denoted respectively as $m\_{s}$, $d\_{s}$, and $k\_{s}$ in Table~\ref{tab:NREL-5MW} and the tower's natural frequency is denoted $\omega\_{n,s}$.

\begin{figure}[t!]
    \centering
    \includegraphics[width=\linewidth]{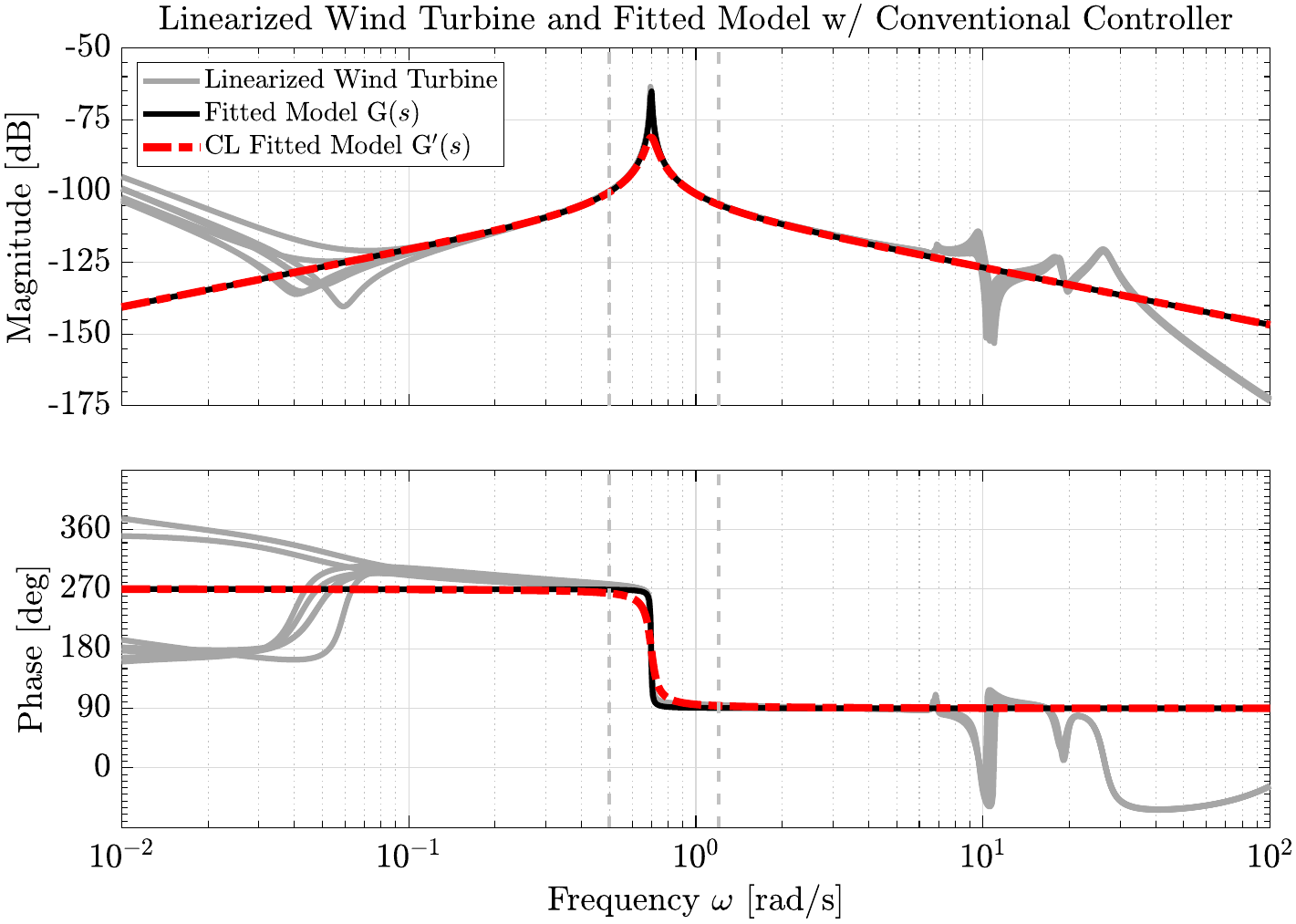}
    \caption{
    Bode plot of linearized NREL 5-MW reference wind turbine and fitted simple model $\mrm{G}(s)$ and closed-loop model with the conventional active tower controller.
    Vertical dashed lines indicate the operating range of the turbine.
    }
    \label{fig:bode_NREL5MW}
\end{figure}

For controller design, the reference wind turbine is linearized at the below-rated wind speeds, ranging from ${v = 4}$~m/s to ${v = 10}$~m/s with $1$~m/s increment.
Figure~\ref{fig:bode_NREL5MW} shows the Bode plots of the linearized wind turbine (gray lines) for the different operating points, where the transfer from the generator torque to tower velocity is taken.
Also plotted is the second order tower model $\mrm{G}(s)$ (black lines), in which modal properties of the scaled reference wind turbine's tower are employed, as well as gain adjustment to fit the linearized wind turbine plots.
The gain adjustment is made by setting $s\_{f}$ to $1.667$~m$^{-1}$, which from the physical point of view may infer that, for the employed reference turbine, the prismatic beam assumption as used in~\eqref{eq:dynamics_tower} might be inaccurate.
Note that numerical artifacts present in the linearized wind turbine at frequencies lower than $0.01$~rad/s, which makes the use of the fitted model $\mrm{G}(s)$ more convenient.

Additional damping is added to the lightly-damped $\mrm{G}(s)$ by the conventional active tower damper explained in Section~\ref{sec:2_baseline_ctrl}.
The gain of the conventional controller is chosen to be ${K\_{I,conv} = -10000}$, which is equivalent to increasing the modal damping into $d\_{eff,s} = 1.9125\cdot10^4$~Ns/m.
The negative sign of $K\_{I,conv}$ is needed to account for the difference in the coordinate convention used in the simple model and OpenFAST.
In Fig.~\ref{fig:bode_NREL5MW}, the Bode plot of the fitted plant $\mrm{G}(s)$ in closed-loop with this conventional controller is shown by the red, dashed lines and denoted $\mrm{G}'(s)$.

Having a damped tower, the next step is to cascade MDCs on top of the conventional controller.
One needs to be reminded that the conventional controller and MDCs serve different purposes and are fundamentally different in that the former increases the effective damping of the tower structure whilst the latter cancels rotational-speed-driven load at the tower.

\begin{figure}[!t]
    \centering
    \subfloat[]{
        \label{fig:bode_NREL5MW_SISO_Mod_Loop}
        \includegraphics[width=\linewidth]{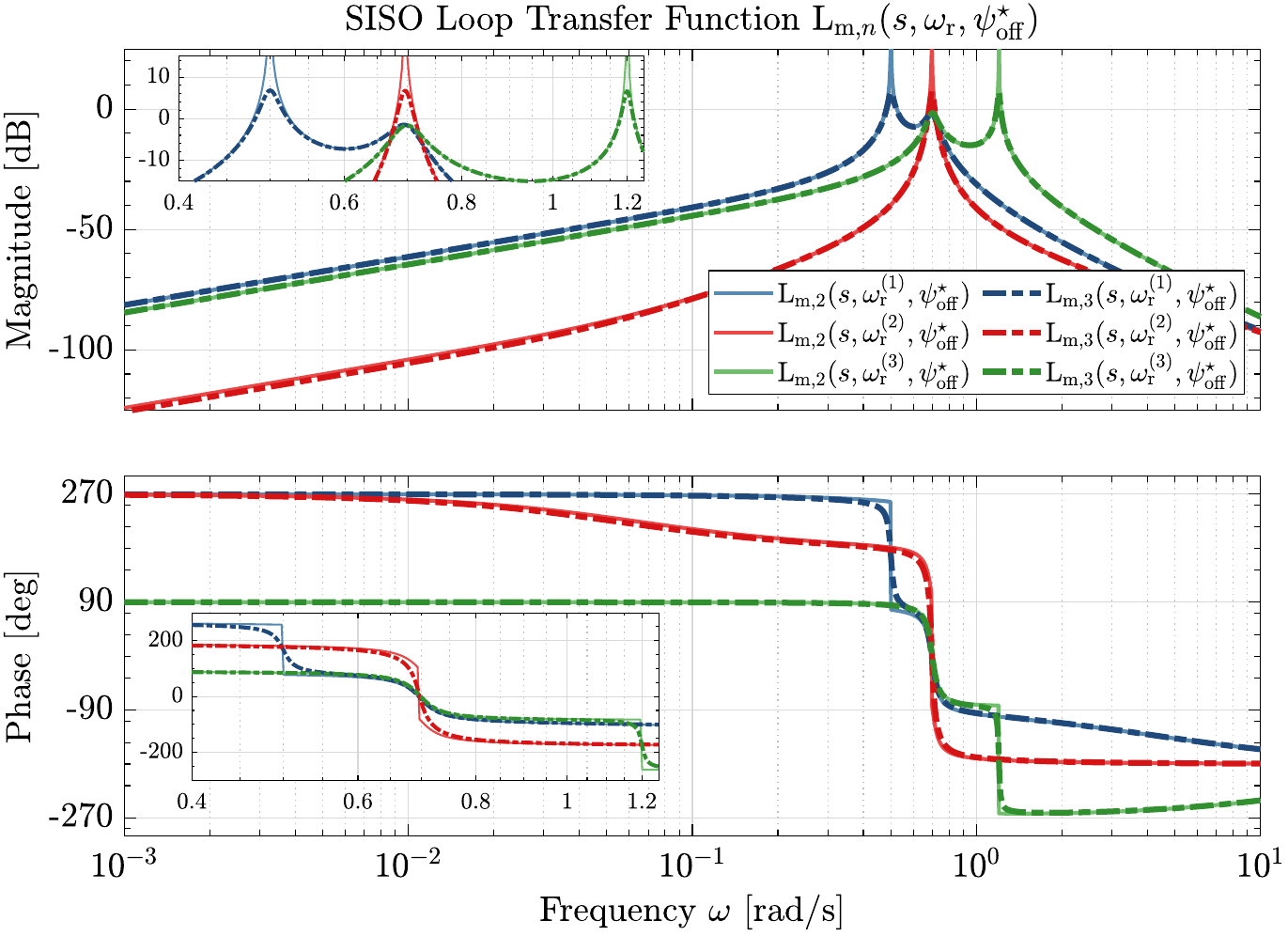}
    }        
    \hfill
    \subfloat[]{
        \label{fig:bode_NREL5MW_MIMO_demod_loop}
        \includegraphics[width=\linewidth]{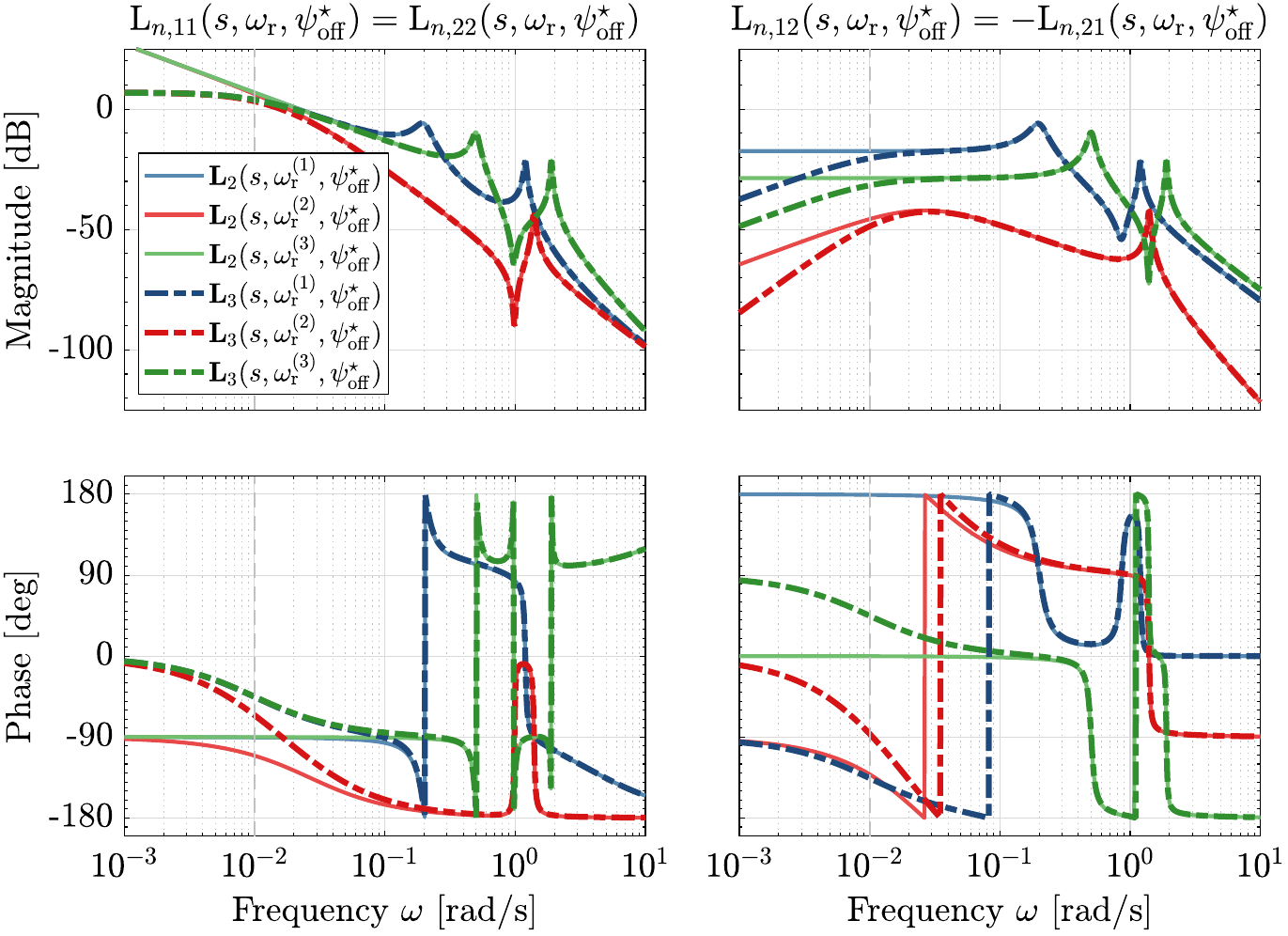}
    }
    \caption{
    Bode plots of SISO and MIMO~ loop transfer functions for ${\omega\_{r}^{(i)}=\{\omega\_{r,min},\omega\_{n,s},\omega\_{r,rated}\}}$.
    In~(a), SISO loop transfers with an undamped inverted notch filter and a damped inverted notch filter are depicted with the respective solid and dashed lines.
    The corresponding transformed MIMO transfers are shown in~(b), where the solid lines illustrate the MIMO demodulated plant with an integral controller and dashed lines with a low-pass filter.
    }
    \label{fig:bode_NREL5MW_loops}
\end{figure}

Similar to the simple wind turbine simulations, the (damped) inverted notch filters $\mrm{C}_{\mrm{m},2}(s,\omega\_{r},\psi\_{off}^\star)$ and $\mrm{C}_{\mrm{m},3}(s,\omega\_{r},\psi\_{off}^\star)$, with ${\omega\_{LPF} = 0.01}$~rad/s used in the latter, are employed.
To cast similar 1P load-reducing performance of MDCs for the entire operating regime, gain-scheduling is implemented by setting the controller gains ${K\_{I} = K\_{L} = 0.022 \gamma}$ with the inverse of the plant's magnitude at the disturbance frequency ${\gamma = 1/|\mrm{G}'(\rmj\omega\_{r})|}$~\cite{Hendrickson2012}.
Figure~\ref{fig:bode_NREL5MW_loops} depicts the resulting Bode plots of the SISO (Fig.~\ref{fig:bode_NREL5MW_SISO_Mod_Loop}) and MIMO loop transfer functions (Fig.~\ref{fig:bode_NREL5MW_MIMO_demod_loop})
\begin{equation*}
    \mrm{L}_{\mrm{m},n}(s,\omega\_{r}^{(i)},\psi\_{off}^\star) = \mrm{G}(s)\mrm{C}_{\mrm{m},n}(s,\omega\_{r}^{(i)},\psi\_{off}^\star) \,,
\end{equation*}
and
\begin{equation*}
    \mbf{L}_{n}(s,\omega\_{r}^{(i)},\psi\_{off}^\star) = \mbf{G}_{2}(s,\omega\_{r}^{(i)},\psi\_{off}^\star)\mbf{C}_{n}(s) \,,
\end{equation*}
respectively, where ${\omega\_{r}^{(i)}=\{\omega\_{r,min},\omega\_{n,s},\omega\_{r,rated}\}}$.
During the constant and turbulent wind cases that follow, the information of $\gamma$ and ${\psi\_{off}^\star=\angle \mrm{G}'(\rmj\omega\_{r})}$ are fed into the MDCs by LUTs, making use of low-pass-filtered rotor speed measurements with the same cut-off frequency as used in the simple wind turbine simulations.

\subsubsection{Constant Wind Simulations}
\noindent
Steady, uniform constant wind cases at $v=\{5,6.25,10\}$~m/s, each lasting for $1000$~s, are employed for simulations in this section, which produce rotor speeds being lower, equal, and higher than the resonance frequency at steady-state.
To induce a 1P excitation to the fixed structure by a rotor mass imbalance, the blade mass properties are modified such that two blades are equally heavier than one other blade.

\begin{figure}[!t]
    \centering
    \includegraphics[width=\linewidth]{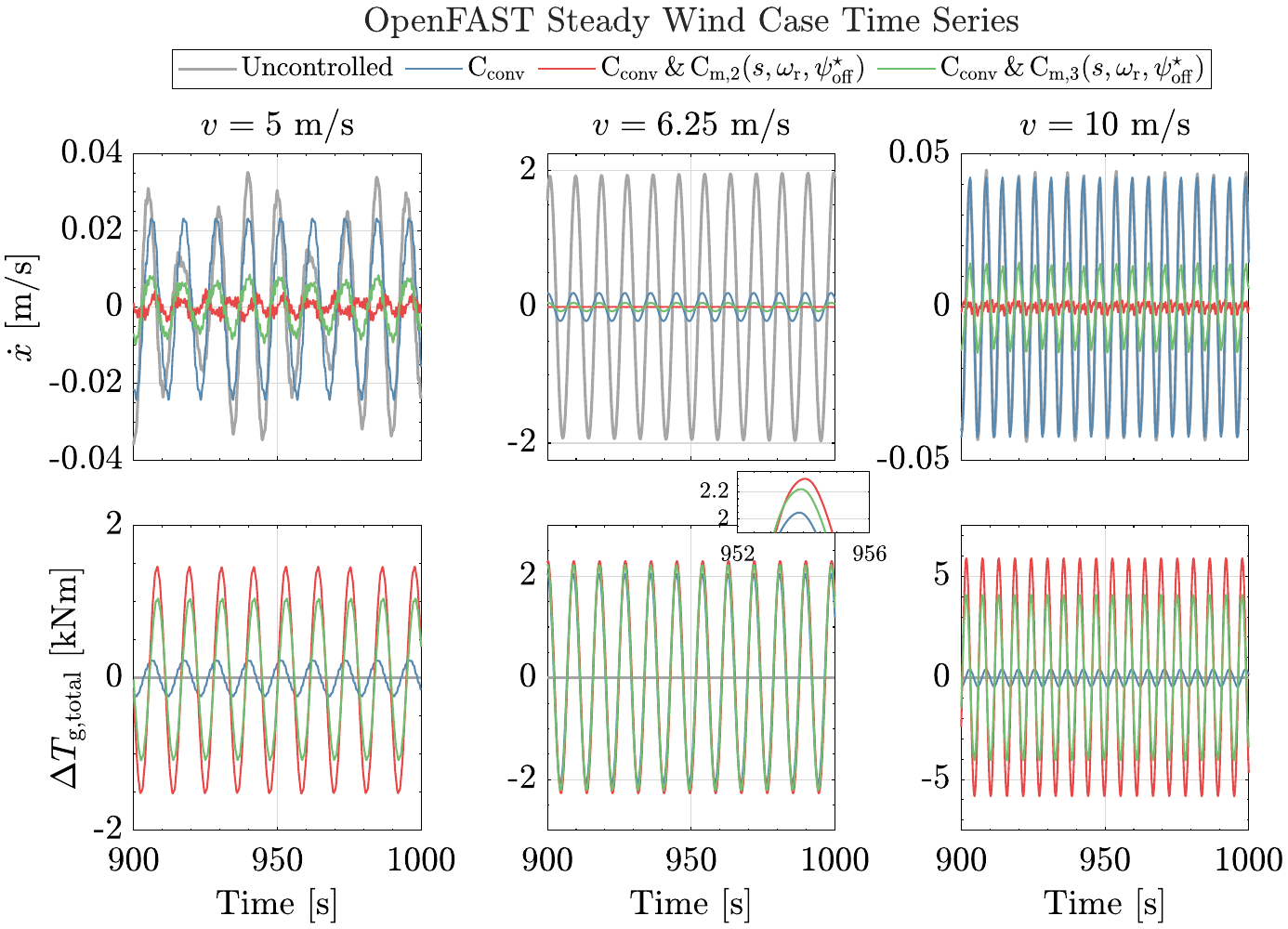}
    \caption{
    OpenFAST time series results in constant wind cases $v=\{5, 6.25,10\}$~m/s.
    During the steady-state at ${t=900-1000}$~s, a large portion of the tower load is mitigated by the conventional controller cascaded with MDCs where $\mrm{C}_{\mrm{m},2}(s,\omega\_{r},\psi\_{off}^\star)$ yields the most reduction.
    Increased controller input at higher wind speeds is caused by the greater amplitude of the 1P periodic load.
    }
    \label{fig:TimeSeries_OpenFAST}
\end{figure}

Figure~\ref{fig:TimeSeries_OpenFAST} shows the wind turbine tower velocity and total additive generator torque measurements ${\Delta T\_{g,total} = \Delta T\_{g,damp}+\Delta T\_{g}}$ for these cases, shown during the steady-state at $t = 900-1000$~s.
The uncontrolled wind turbine signals are shown by the gray lines and those with only the conventional controller $\mrm{C}\_{conv} = K\_{conv}$ are shown by the blue lines.
The conventional tower damper targets fatigue loading at the tower's natural frequency, while tower excitations at other frequencies are not alleviated as effectively, especially the 1P-driven load.
This is evident in the figure, where the conventional controller performs well only in the second steady wind case where $\omega\_{r}=\omega\_{n}$.

Nevertheless, some residual oscillations are still shown.
The performance of $\mrm{C}\_{conv}$ is improved by cascading it with the MDCs $\mrm{C}_{\mrm{m},2}(s,\omega\_{r},\psi\_{off}^\star)$ and $\mrm{C}_{\mrm{m},3}(s,\omega\_{r},\psi\_{off}^\star)$, illustrated by the respective red and green lines.
Similar to the simplified wind turbine simulations in Section~\ref{sec:6_simple_turbine}, the infinite gain of $\mrm{C}_{\mrm{m},2}(s,\omega\_{r},\psi\_{off}^\star)$ at the 1P frequency creates the most control effort in every case compared to other settings.
This consequently allows the controller to mitigate most of the periodic loads while still providing damping at the tower's natural frequency.
Less 1P load reduction due to the less aggressive control action at this frequency is demonstrated when $\mrm{C}\_{conv}$ is combined with $\mrm{C}_{\mrm{m},3}(s,\omega\_{r},\psi\_{off}^\star)$, while still outperforming $\mrm{C}\_{conv}$ without MDCs.
Also noticeably different than the simplified wind turbine simulations is the effect of the gain-scheduling of the MDCs, such that the effectiveness of the 1P load reduction is not only observed during the resonance but also when $\omega\_{r}<\omega\_{n}$ at $v=5$~m/s and $\omega\_{r}>\omega\_{n}$ at $v=10$~m/s.

\begin{rem}
The rotor mass imbalance creates greater centrifugal force when the rotor spins faster, resulting in greater 1P loading amplitude at higher rotational speeds, in contrast to the constant amplitude assumed in simplified wind turbine simulations.
This explains the need for larger control action of the cascaded controllers for higher rotor speeds.
\end{rem}

\subsubsection{Turbulent Wind Simulations}
\noindent
Two different turbulent cases are chosen with mean wind speed at hub-height ${v\_{h}=6.25}$~m/s, namely, ${I\_{T}=4\%}$ and ${I\_{T}=12\%}$, representing low and high turbulence, respectively.
For these wind cases, $2000$~s of simulations are run, where the first $200$~s is not accounted for to exclude transient effects from the analysis.
The same rotor mass imbalance from the steady wind simulations is used here to induce 1P loading.

\begin{figure}[!t]
    \centering
    \subfloat[]{
        \label{fig:TimeSeries_OpenFAST_Turbs}
        \includegraphics[width=\linewidth]{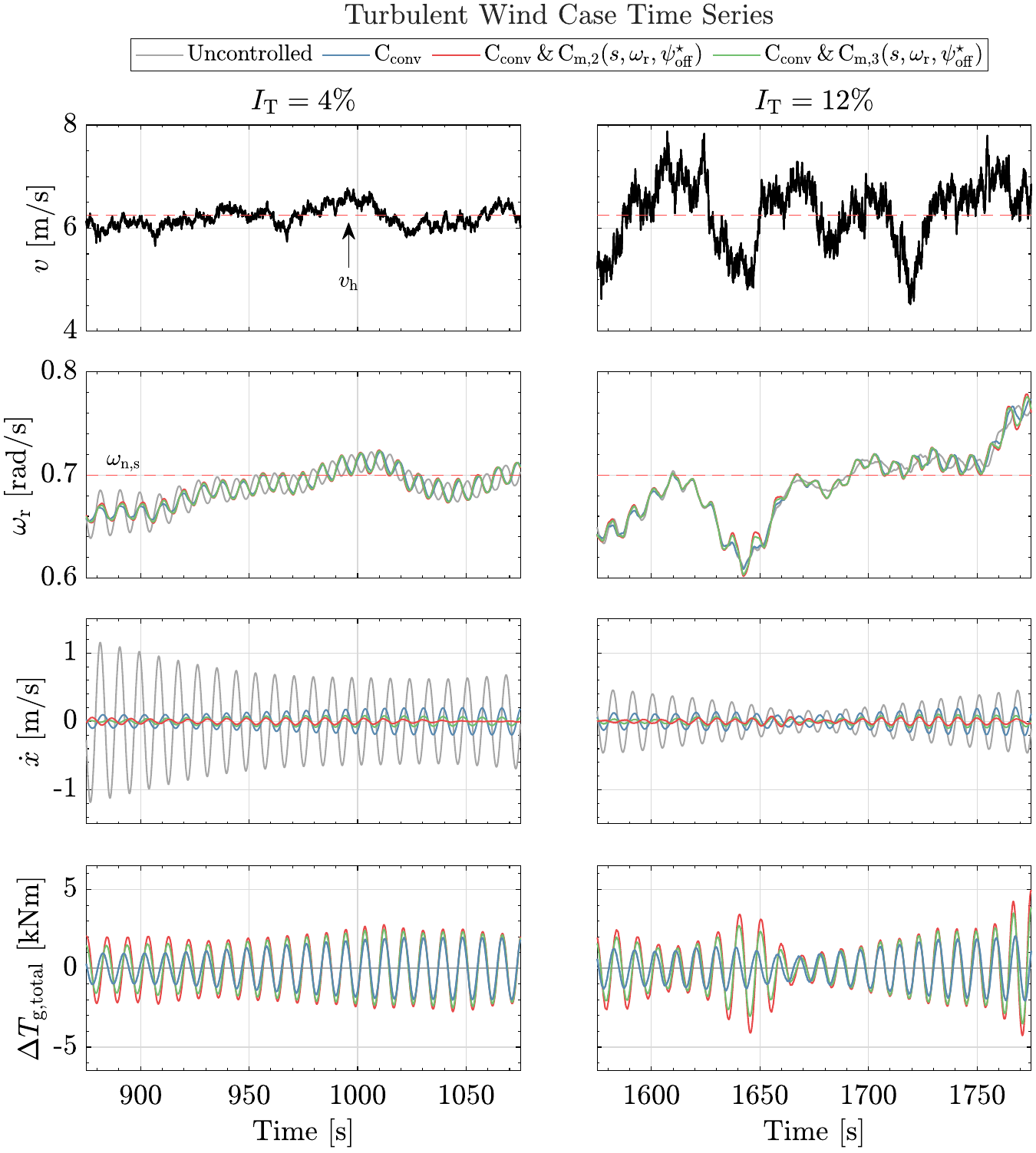}
    }        
    \hfill
    \subfloat[]{
        \label{fig:PSD_OpenFAST_Turbs}
        \includegraphics[width=\linewidth]{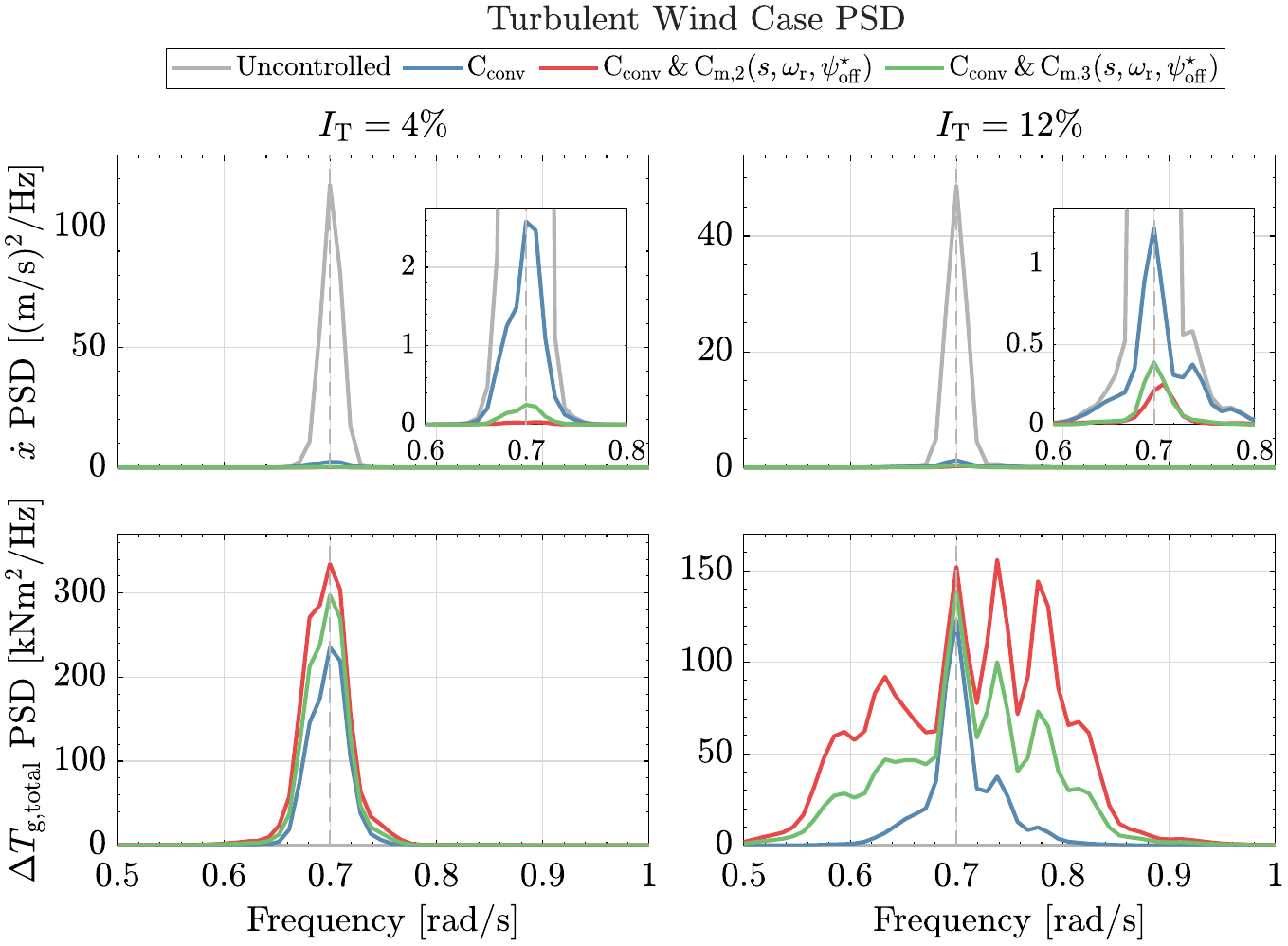}
    }
    \caption{
    Time series (a) and PSD (b) results of the turbulent wind cases.
    Cascaded conventional tower damper and MDCs outperform the conventional controller without MDCs in both low and high turbulence.
    For ${v\_{h}=6.25}$~m/s, different turbulence intensities influence the prevalence of resonances, as shown in (a), thereby affecting the PSD content about ${\omega\_{n,s}=0.6963}$~rad/s (gray dashed lines) in (b).
    Greater variation in the 1P frequency during high turbulence results in the cascaded controllers actively operating in a wider frequency range to mitigate the periodic load, shown clearly in the PSD of $\Delta T\_{g,total}$.
    }
    \label{fig:OpenFAST_Turbs}
\end{figure}

Results of these turbulent cases are presented in Fig.~\ref{fig:OpenFAST_Turbs}.
Figure~\ref{fig:TimeSeries_OpenFAST_Turbs} depicts the time series results, where records at $t=875-1075$~s and $t=1575-1775$~s are shown for the respective low and high turbulence cases.
In the figure, $v$, $\omega\_{r}$, $\dot{x}$, and $\Delta T\_{g,total}$ measurements are shown from the first to the fourth rows, respectively.
Also indicated by the red dashed lines in the measurements of $v$ and $\omega\_{r}$ are $v\_{h}$ and $\omega\_{n}$.
Figure~\ref{fig:PSD_OpenFAST_Turbs} depicts the corresponding power spectral density (PSD) plots of $\dot{x}$ and $\Delta T\_{g,total}$, post-processed from the time series measurements.
In low turbulence where $v$ varies closer to $v\_{h}$, $\omega\_{r}$ tends to cause tower resonance more frequently compared to when turbulence is higher.
This explains the higher PSD magnitude of $\dot{x}$ about $\omega\_{n}$ (gray dashed lines) during low turbulence with respect to its higher turbulent counterpart.
Regarding controllers' activity, results consistent with the previous steady wind simulations are observed in the two turbulence cases.
In the time series, $\mrm{C}\_{conv}$ is shown to perform less effectively than when operated in conjunction with MDCs.
In terms of load reduction, $\mrm{C}\_{conv}$ with $\mrm{C}_{\mrm{m},2}(s,\omega\_{r},\psi\_{off}^\star)$ performs best.
More benign control action is exercised when $\mrm{C}_{\mrm{m},3}(s,\omega\_{r},\psi\_{off}^\star)$ is incorporated but resultingly, slightly less reduction in tower fatigue load is performed.

The capabilities of the MDCs to follow and cancel the varying 1P periodic load frequency are shown best by the high turbulence case, where 1P frequency varies more and covers a wider range than in low turbulence.
This is most evident in the PSD result of $\dot{x}$, lower frequency content with respect to $\mrm{C}\_{conv}$ is evident not only at $\omega\_{n}$ but also at the surrounding frequencies.
In the PSD plot of $\Delta T\_{g,total}$ in Fig.~\ref{fig:PSD_OpenFAST_Turbs}, multiple peaks at $0.6\,,0.63\,,0.74$, and $0.78$~rad/s are seen for the cascaded controller settings, apart from that at $\omega\_{n}$.
This indicates intensive 1P load reduction activity at the said frequencies, which are virtually non-existent except at $0.74$~rad/s for the $\mrm{C}\_{conv}$ setting.

A statistical evaluation of the measurement data from the simulations is done in terms of standard deviations.
Table~\ref{tab:OpenFAST_results} summarizes those of the side-side tower velocity $\sigma_{\dot{x}}$, total additive generator torque $\sigma_{\Delta T\_{g,total}}$, and generated power $\sigma_{P\_{g}}$.
Respectively, these values indicate changes in the structural fatigue load, controller activity, and power fluctuation.
Upward ($\uparrow$) and downward ($\downarrow$) arrows are used to indicate standard deviations that are higher and lower with respect to the uncontrolled turbine.
The overall computed $\sigma_{\dot{x}}$ and $\sigma_{\Delta T\_{g,total}}$ for both turbulent cases point to the same conclusion as the time series and PSD results.
Best fatigue load reduction, shown by the least $\sigma_{\dot{x}}$ values, is achieved by $\mrm{C}\_{conv}$ and $\mrm{C}_{\mrm{m},2}(s,\omega\_{r},\psi\_{off}^\star)$, the control action of which is also the most active, as indicated by the corresponding $\sigma_{\Delta T\_{g,total}}$.
With respect to this configuration, $\mrm{C}\_{conv}$ and $\mrm{C}_{\mrm{m},3}(s,\omega\_{r},\psi\_{off}^\star)$ are able to compromise between the actuation effort and load mitigation, shown by their milder $\sigma_{\Delta T\_{g,total}}$ from which slightly lower $\sigma_{\dot{x}}$ is obtained.
Again, $\mrm{C}\_{conv}$ shows the least increase in $\sigma_{\Delta T\_{g,total}}$ with respect to the uncontrolled case, but also the least load reduction among other controller setups.
As torque fluctuation affects the generator power due to their proportional relation, the trend of $\sigma_{P\_{g}}$ follows that of $\sigma_{\Delta T\_{g,total}}$.
Important to note here is during the high turbulence, generated power fluctuates more due to the high variation in the wind, thus the much higher overall standard deviation with respect to the low turbulence case.
Improvements in terms of power fluctuation may be achieved by utilization of individual blade pitching in place of additive generator torque due to the less coupling with the generator power~\cite{Pamososuryo2022b}.
Having the simulation results analyzed, the conclusions of this work are drawn in the next section.

\begin{table}[!t]
    \caption{
    Standard deviations of tower velocity, total additive generator torque, and generated power for the turbulent wind cases.
    }
    \centering
    \begin{tabular}{llrrr}
        \hline
        \begin{tabular}[c]{@{}r@{}}$I\_{T}$\\(\%)\end{tabular} & \textbf{Controller} & \begin{tabular}[c]{@{}r@{}}$\sigma_{\dot{x}}$\\(m/s)\end{tabular} & \begin{tabular}[c]{@{}r@{}}$\sigma_{\Delta T\_{g,total}}$\\(kNm)\end{tabular} & \begin{tabular}[c]{@{}r@{}}$\sigma_{P\_{g}}$\\(kW)\end{tabular} \\ [0.5ex]
        \hline\hline
        \multirow{4}{*}{$4$}  & Uncontrolled                                                                   & $0.646$           & $0$             & $67.924$ \\
                              & $\mrm{C}\_{conv}$                                                              & $\downarrow0.126$ & $\uparrow1.263$ & $\uparrow113.249$ \\
                              & $\mrm{C}\_{conv}\,\&\,\mrm{C}_{\mathrm{m},2}(s,\omega\_{r},\psi\_{off}^\star)$ & $\downarrow0.016$ & $\uparrow1.638$ & $\uparrow132.500$ \\
                              & $\mrm{C}\_{conv}\,\&\,\mrm{C}_{\mathrm{m},3}(s,\omega\_{r},\psi\_{off}^\star)$ & $\downarrow0.041$ & $\uparrow1.476$ & $\uparrow123.685$ \\
        [1ex] 
        \hline
        \multirow{4}{*}{$12$} & Uncontrolled                                                                   & $0.393$           & $0$             & $201.579$ \\
                              & $\mrm{C}\_{conv}$                                                              & $\downarrow0.095$ & $\uparrow0.945$ & $\uparrow212.809$ \\
                              & $\mrm{C}\_{conv}\,\&\,\mrm{C}_{\mathrm{m},2}(s,\omega\_{r},\psi\_{off}^\star)$ & $\downarrow0.041$ & $\uparrow1.953$ & $\uparrow245.641$ \\
                              & $\mrm{C}\_{conv}\,\&\,\mrm{C}_{\mathrm{m},3}(s,\omega\_{r},\psi\_{off}^\star)$ & $\downarrow0.047$ & $\uparrow1.525$ & $\uparrow229.328$ \\
        [1ex] 
        \hline
        \multicolumn{5}{l}{$\uparrow/\downarrow$: increase/decrease w.r.t. uncontrolled wind turbine.}
    \end{tabular}
    \label{tab:OpenFAST_results}
\end{table}

\section{Conclusions}\label{sec:7}   
\noindent
In this paper, an MDC framework for the cancellation of 1P periodic loading acting on wind turbine side-side tower motion has been proposed.
The framework relies on the modulation of input and demodulation of output signals at the periodic load frequency, resulting in each signal being representable in its quadrature and in-phase components.
Convenient yet effective diagonal LTI controllers are designed onto these channels, representable as an LTV when combined with the modulation-demodulation.
MIMO representation of the plant has also been rendered in terms of the quadrature and in-phase channels, which, by frequency-domain analysis, has been shown to contain cross-coupling at steady-state and instability-inducing gain sign flip.
A phase offset, the optimal value of which is defined as the negative of the nominal plant's phase at the 1P frequency, has been shown to be a remedy for both the cross-coupling and gain sign flip issues.
Simulations at two different levels of fidelity have been conducted to demonstrate the effectiveness of two proposed MDC designs, being undamped and damped inverted notch filters centered at the 1P frequency.
Low-fidelity simulations exhibited the controllers' performance deterioration and instability when the optimal phase offset was not incorporated.
OpenFAST was employed to simulate steady and turbulent wind cases in a higher-fidelity setting, in which the MDCs are cascaded with a conventional tower damping controller.
Results have indicated a performance improvement of the conventional controller in fatigue load reduction when the MDCs are operated synergetically.
The conventional tower damper has been shown to mitigate only the tower's natural frequency, while the MDCs target the 1P periodic loading caused by a mass imbalance in the rotor disk.



\bibliographystyle{ieeetr}
\bibliography{references.bib}

\begin{IEEEbiography}[{\includegraphics[width=1in,height=1.25in,clip,keepaspectratio]{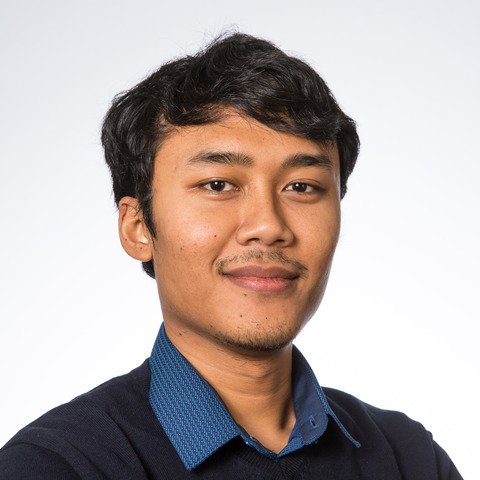}}]{Atindriyo K. Pamososuryo} was born in Medan, Indonesia, in 1991. 
He received his M.Sc. degree from the Delft Center for Systems and Control, Delft University of Technology, the Netherlands, in 2018, where he is currently pursuing his Ph.D.
His research interests include, but are not limited to, dynamical system modeling, state estimation, linear parameter varying systems, and model predictive control. 
In particular, his main focus is on the applications in the wind energy field for advancements in load reduction and power production capabilities at the wind turbine level, in which he also collaborated with Vestas Wind Systems A/S.
In 2023, he received the O. Howard Schuck Award from the American Automatic Control Council for his contribution to the applied control engineering field.
\end{IEEEbiography}
\vskip 0pt plus -1fil
\begin{IEEEbiography}[{\includegraphics[width=1in,height=1.25in,clip,keepaspectratio]{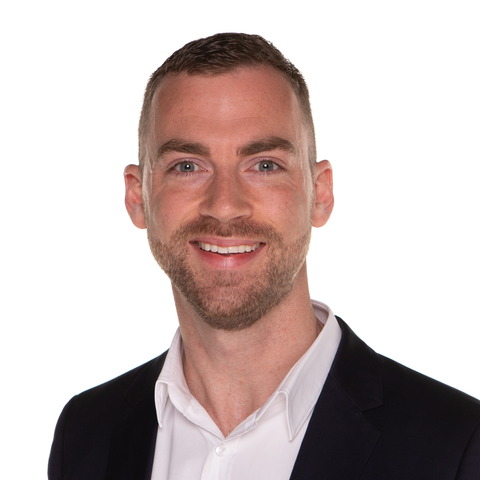}}]{Sebastiaan P. Mulders}
received his M.Sc. Systems and Control in 2015, and his Ph.D. in wind turbine control in 2020 at the Delft University of Technology, The Netherlands. Since 2021 he has been a postdoctoral researcher on the development of learning algorithms for wind turbine controllers with the Delft University of Technology, where he has been an assistant professor since 2022. His main research interest is in data-enabled control co-design. The co-design approach is supported by machine learning techniques for the synergetic and efficient optimization of the system and controller to reach higher performance levels for present-day complex systems with increased levels of interactions.
\end{IEEEbiography}
\vskip 0pt plus -1fil
\begin{IEEEbiography}[{\includegraphics[width=1in,height=1.25in,clip,keepaspectratio]{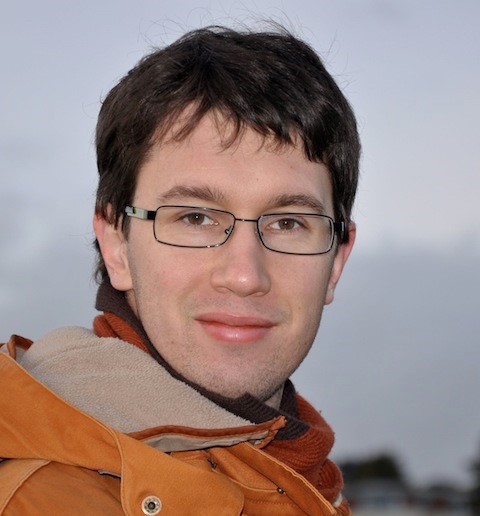}}]{Riccardo Ferrari}
received the Laurea degree with and the Ph.D. degree from University of Trieste, Italy. He held both academic and industrial R\&D positions, in particular as researcher in the field of process instrumentation and control for the steel-making sector. He is a Marie Curie alumnus and currently an Associate Professor with the Delft Center for Systems and Control, Delft University of Technology, The Netherlands. His research interests include fault tolerant control and fault diagnosis and attack detection in large-scale cyber–physical systems, with applications to wind energy generation, electric mobility, and cooperative autonomous vehicles.
\end{IEEEbiography}
\vskip 0pt plus -1fil
\begin{IEEEbiography}[{\includegraphics[width=1in,height=1.25in,clip,keepaspectratio]{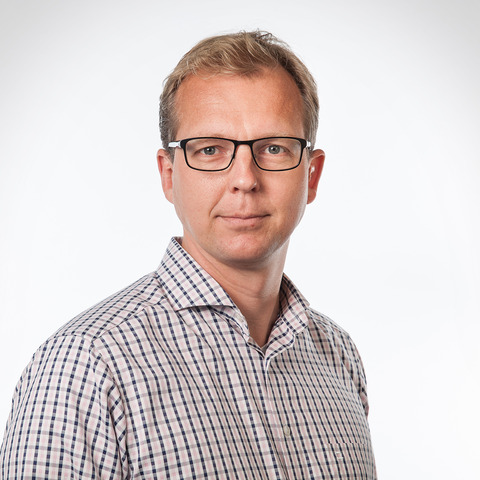}}]{Jan-Willem van Wingerden}
(Senior Member, IEEE) was born in Ridderkerk, The Netherlands, in 1980. He received the B.S. and Ph.D. (cum laude) degrees in mechanical engineering and in control engineering from the Delft Center for Systems and Control, Delft University of Technology, Delft, The Netherlands, in 2004 and 2008, respectively. His Ph.D. thesis was entitled Smart Dynamic Rotor Control for Large Offshore Wind Turbines. He is currently a full Professor with the Delft University of Technology. His current research is mainly centered around the development of data driven controllers for wind turbines and wind farms.
\end{IEEEbiography}

\end{document}